\documentclass[aps,prx,twocolumn,reprint]{revtex4-2}

\usepackage{graphicx}% Include figure files
\usepackage{dcolumn}% Align table columns on decimal point
\usepackage{bm}% bold math
\usepackage{amsmath,amssymb,mathtools,braket}
\usepackage{color,comment,footnote}

\usepackage{amsthm}
\theoremstyle{plain}
\newtheorem{thm}{Theorem}

\usepackage{hyperref}% add hypertext capabilities
\usepackage{xcolor}
\hypersetup{
	colorlinks=true,
	citecolor=blue,
	linkcolor=blue,
	urlcolor=blue,
}

\newcommand{\mO}{\mathcal{O}}

\newcommand{\tr}{\text{tr}}

\begin{document}
	
	%\preprint{APS/123-QED}
	
	\title{Splitting and Parallelizing of Quantum Convolutional Neural Networks for Learning Translationally Symmetric Data}% Force line breaks with \\

	\author{Koki Chinzei}\thanks{chinzei.koki@fujitsu.com}
	\author{Quoc Hoan Tran}
	\author{Kazunori Maruyama}
	\author{Hirotaka Oshima}
	\author{Shintaro Sato}
	\affiliation{Quantum Laboratory, Fujitsu Research, Fujitsu Limited,
		4-1-1 Kawasaki, Kanagawa 211-8588, Japan}

	\date{\today}% It is always \today, today,

	\begin{abstract}
		
		The quantum convolutional neural network (QCNN) is a promising quantum machine learning (QML) model that is expected to achieve quantum advantages in classically intractable problems.
		However, the QCNN requires a large number of measurements for data learning, limiting its practical applications in large-scale problems.
		To alleviate this requirement, we propose a novel architecture called split-parallelizing QCNN (sp-QCNN), which exploits the prior knowledge of quantum data to design an efficient model.
		This architecture draws inspiration from geometric quantum machine learning and targets translationally symmetric quantum data commonly encountered in physics and quantum computing science.
		By splitting the quantum circuit based on translational symmetry, the sp-QCNN can substantially parallelize the conventional QCNN without increasing the number of qubits and improve the measurement efficiency by an order of the number of qubits.
		To demonstrate its effectiveness, we apply the sp-QCNN to a quantum phase recognition task and show that it can achieve comparable classification accuracy to the conventional QCNN while considerably reducing the measurement resources required. 
		Due to its high measurement efficiency, the sp-QCNN can mitigate statistical errors in estimating the gradient of the loss function, thereby accelerating the learning process.
		These results open up new possibilities for incorporating the prior data knowledge into the efficient design of QML models, leading to practical quantum advantages.
		
	\end{abstract}
	
	\maketitle

	%%%%%%%%%%%%%%%%%%%%%%%%%%%%%%%%%%%%%%%%%%%%%%%%%%%%%%%%%%%%%%%%%%%%%%%%
	%%%%%%%%%%%%%%%%%%%%%%%%%%%%%%%%%%%%%%%%%%%%%%%%%%%%%%%%%%%%%%%%%%%%%%%%
	%%%%%%%%%%%%%%%%%%%%%%%%%%%%%%%%%%%%%%%%%%%%%%%%%%%%%%%%%%%%%%%%%%%%%%%%
	\section{Introduction}

	%%%%%%%%%%%%%%%%%%%%%%%%%%%%%%%%%%%%%%%%%%%%%%%%%%%%%%%%%%%%%%%%%%%%%%%%
	\begin{figure*}[t]
		\centering
		\includegraphics[width=\linewidth]{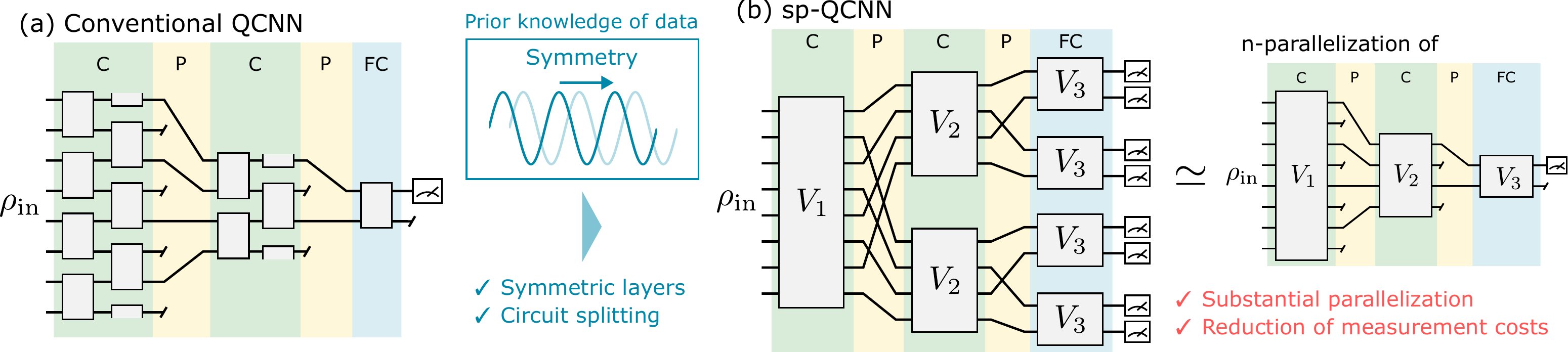}
		\caption{
			Basic structures of (a) conventional and (b) sp-QCNNs.
			In the figure, C, P, and FC represent convolutional, pooling, and fully-connected layers, respectively.
			(a) In the conventional QCNN, some qubits are discarded at each pooling layer, and only one of the remaining qubits is measured in the end to classify the quantum data.
			(b) In the sp-QCNN, the translational symmetry of data is used as prior knowledge to design an efficient QML model. 
			The circuit of the sp-QCNN (the left circuit) consists of translationally symmetric layers and splitting structures, allowing us to substantially parallelize the nonsplitting QCNN (the right circuit) to improve the measurement efficiency.
		}
		\label{fig: illust}
	\end{figure*}
	%%%%%%%%%%%%%%%%%%%%%%%%%%%%%%%%%%%%%%%%%%%%%%%%%%%%%%%%%%%%%%%%%%%%%%%%

	Quantum computing is an innovative technology that is expected to solve classically intractable problems and open up new frontiers in scientific research and technological advancements~\cite{Nielsen2010-pf}.
	Quantum machine learning (QML) is one of the central research fields in quantum computing, allowing us to solve various tasks such as classification, regression, and clustering by discovering relationships and patterns between data using quantum computers~\cite{Wiebe2014-un, Schuld2015-su, Biamonte2017-fz, Dunjko2018-as}.
	Recent studies have demonstrated quantum speedups in QML beyond classical machine learning for specific artificially engineered tasks, suggesting the potential of QML~\cite{Havlicek2019-wd, Liu2021-oe, Huang2021-wb}. 
	While many studies have been devoted to achieving quantum advantages of QML for practical problems, the path to achieving this goal remains unclear~\cite{Schuld2022-wb}.

	The quantum neural network (QNN) is a promising QML model that combines the principles of quantum information processing and artificial neural networks to enhance the capabilities of data-driven technologies~\cite{Farhi2018-nt, Liu2018-bd, Mitarai2018-ap, Benedetti2019-ph, Schuld2020-vz}. 
	The QNN is represented by a parametrized quantum circuit, which is optimized via training data to solve a given task~\cite{Cerezo2021-un}.
	Since the efficient simulation of quantum circuits is generally impossible with classical computers, the QNN can learn the complex features of data that are classically intractable~\cite{Du2020-ty, Abbas2021-mh}. 
	Among various QNN architectures, the quantum convolutional neural network (QCNN) is a leading one that enables classification tasks~\cite{Grant2018-re, Cong2019-ov} [Fig.~\ref{fig: illust}(a)].
	For instance, the QCNN can classify the phases of matter in quantum many-body systems, an important research object in the broad field of physics~\cite{Herrmann2022-bs, Liu2022-uu, Monaco2023-xr}.
	Due to its high trainability and feasibility~\cite{Pesah2021-py}, the QCNN is particularly suitable for noisy intermediate-scale quantum (NISQ) devices with a limited number of possible gate operations~\cite{Preskill2018-xr}.

	It is conjectured that achieving quantum advantages in QML requires encoding the prior knowledge of a problem, or inductive bias, into the learning models~\cite{Kubler2021-qx}.  
	In QNNs, the architecture design tailored to prior knowledge is considered crucial to take full advantage of its capabilities~\cite{Cerezo2022-xy}.
	Geometric quantum machine learning (GQML) based on equivariant QNNs, where the symmetry of a problem is encoded in a variational unitary circuit, is one of such prior knowledge-tailored learning models, reducing the parameter space to be searched and enhancing trainability and generalization~\cite{Bronstein2021-is, Verdon2019-pi, Zheng2023-ze, Larocca2022-mj, Meyer2023-vx, Skolik2022-ge, Sauvage2024-mv, Ragone2022-va, Nguyen2022-go}.
	For example, it was theoretically proved that permutation-equivariant QNNs~\cite{Schatzki2022-or} do not suffer from barren plateaus~\cite{McClean2018-qf, Cerezo2021-tq, Ortiz_Marrero2021-ni, Holmes2022-uk}, the exponential vanishing of the gradient in a loss function, due to the exponential reduction of parameters to reach overparametrization~\cite{Larocca2021-ti}.
	This technique would be a powerful tool to achieve quantum advantages with QNNs.

	The high resource requirement of the measurement process remains a practical barrier for QNNs to learn data on real quantum computers~\cite{Cerezo2021-un}. 
	During the learning process of QNNs, a predefined loss function is minimized by adjusting the variational parameters of the circuit.
	This loss function is computed from a training dataset by measuring specific observables in the parameterized quantum circuit.
	Therefore, the measurement cost scales with the number of parameters to be optimized and the amount of data to be processed~\cite{Abbas2023-oo}.
	This situation presents a significant bottleneck when considering large-scale QML applications and the potential of practical quantum advantages~\cite{Schuld2022-wb, Liu2023-pv}. 
	To mitigate this measurement requirement, one possible solution is the multiprogramming of quantum computation, which allows multiple circuits to be executed in parallel on different regions of a quantum processor~\cite{Das2019-yw, Liu2021-eo, Niu2021-mv, Niu2023-ki}. 
	Although this parallelization reduces the total runtime, it increases the required qubit resources, which are limited in current devices.

	We address this issue by proposing a novel QNN architecture called split-parallelizing QCNN (sp-QCNN).
	This architecture is inspired by GQML and targets translationally symmetric data, such as solid-state materials in condensed matter physics.
	This model exploits the data symmetry as prior knowledge to substantially parallelize the QCNN without increasing the number of qubits, improving the measurement efficiency [Fig.~\ref{fig: illust}(b)].
	The circuit of the sp-QCNN consists of two elements: translationally symmetric layers and circuit splitting.
	First, we impose translational symmetry on the convolutional and fully-connected layers to preserve the symmetry of the input state.
	Second, we split the circuit (rather than discarding some qubits) at the pooling layers and then perform the same unitary operations on each branch in parallel.
	The combination of this circuit structure and data symmetry substantially parallelizes the conventional QCNN consisting of the same unitary layers, improving the measurement efficiency of local observables and their gradients by a factor of $\mO(n)$ ($n$ is the number of qubits throughout this paper).

	For verification, we apply the sp-QCNN to a quantum phase recognition task.
	The results show that the sp-QCNN can improve the measurement efficiency by a factor of $\mO(n)$ while achieving sufficient classification performance to recognize the symmetry-protected topological (SPT) phase~\cite{Gu2009-zd, Pollmann2010-ed, Chen2011-yh, Pollmann2012-lg}.
	In training with limited measurement resources, the sp-QCNN with high measurement efficiency can suppress statistical errors in estimating the gradient of the loss function to accelerate the learning process compared to the conventional QCNN.
	Our model opens up a new research direction in the QNN architecture design, contributing to practical quantum advantages in near-term quantum devices that lack sufficient computational resources.

	The remainder of this paper is organized as follows.
	First, Sec.~\ref{sec: review} briefly reviews the QCNN and discusses its computational cost.
	Section~\ref{sec: sp-QCNN} introduces two key components of the sp-QCNN,  translationally symmetric layers and circuit splitting, and clarifies the similarities and differences between the sp-QCNN and the GQML.
	Section~\ref{sec: mechanism} shows the advantage of the sp-QCNN, i.e., the improvement of measurement efficiency for local observables and their gradients, based on symmetry.
	For verification, Sec.~\ref{sec: app} presents the application of the sp-QCNN to a quantum phase recognition task, showing that it can solve the task with sufficient accuracy and improve the measurement efficiency by a factor of $\mO(n)$.
	Finally, Sec.~\ref{sec: conclusion} summarizes this paper and discusses potential future research directions.

	%%%%%%%%%%%%%%%%%%%%%%%%%%%%%%%%%%%%%%%%%%%%%%%%%%%%%%%%%%%%%%%%%%%%%%%%
	%%%%%%%%%%%%%%%%%%%%%%%%%%%%%%%%%%%%%%%%%%%%%%%%%%%%%%%%%%%%%%%%%%%%%%%%
	%%%%%%%%%%%%%%%%%%%%%%%%%%%%%%%%%%%%%%%%%%%%%%%%%%%%%%%%%%%%%%%%%%%%%%%%
	\section{Review of QCNN} \label{sec: review}

	The convolutional neural network (CNN) is a celebrated classical machine learning model that solves various tasks, such as image classification~\cite{LeCun2015-yl, Lecun1995-ot, Krizhevsky2012-kx}.
	The CNN consists of three different types of layers: convolutional, pooling, and fully connected layers.
	The convolutional layer filters the input data to extract its local features, and the pooling layer coarse-grains the data to leave only relevant information.
	After the convolutional and pooling layers are alternately applied, the fully connected transformation is applied to the remaining data to produce a final output.
	For example, in classification problems, the output indicates which class the input data belongs to, and the CNN is trained to correctly classify training data.

	The QCNN is a CNN-inspired QNN model that can process quantum data whose dimension is exponentially larger than classical ones and is expected to achieve practical quantum advantages~\cite{Grant2018-re, Cong2019-ov}.
	Similar to the CNN, the QCNN consists of convolutional, pooling, and fully connected layers [Fig.~\ref{fig: illust}(a)].
	The convolutional layers apply local unitary gates to extract the local features of the input data, and the pooling layers discard some qubits to coarse-grain the quantum information.
	After alternately applying the two types of layers, we perform the fully connected unitary, measure the remaining qubits, and obtain an output indicating the data class.
	In the QCNN, the quantum circuit is characterized by variational parameters, which are optimized to correctly classify training data.
	Such a variational algorithm is central in the NISQ era, as it works even in a relatively shallow circuit~\cite{Preskill2018-xr}.

	The QCNN is promising for quantum advantages in NISQ devices because of its two significant features.
	One is its high feasibility.
	Since the number of qubits in the QCNN decreases exponentially in each pooling layer, the circuit depth is $\mO(\log n)$.
	This logarithmic depth is advantageous for NISQ devices where the number of possible gate operations is limited.
	The other feature of the QCNN is its high trainability.
	In many variational quantum algorithms, the exponential vanishing of the gradient in a loss function, known as the barren plateau phenomenon, prevents scalable optimization~\cite{McClean2018-qf, Cerezo2021-tq, Ortiz_Marrero2021-ni, Holmes2022-uk}.
	Meanwhile, Ref.~\cite{Pesah2021-py} proved that the QCNN does not suffer from the barren plateaus due to the logarithmic depth and the locality of unitary operations and observables.
	The absence of barren plateaus leads to the high trainability of the QCNN, which is crucial for achieving quantum advantages in QML tasks.

	However, the high resource requirement of measurements for optimization presents practical difficulties in QNNs, including QCNN~\cite{Cerezo2021-un}.
	Let us estimate the required measurement cost in the QCNN.
	First, we suppose that half of the qubits are discarded at each pooling layer and the number of variational parameters is $\mO(n)+\mO(n/2)+\mO(n/4)+\cdots \sim \mO(n)$ in total
	[in common QCNNs, gates acting in parallel share the same parameters, thus, the number of independent parameters is $\mO(\log n)$, but measuring the gradient of the loss function requires $\mO(n)$ cost (see Sec.~\ref{sec: grad_eff} for details)].
	We also let $N_\text{train}, N_\text{epoch}$, and $N_\text{shot}$ denote the number of training data, maximum epoch (one epoch refers to a complete iteration through a dataset), and the number of measurement shots used per observable, respectively.
	Then, the total required number of shots during training is $\mO(nN_\text{train}N_\text{epoch}N_\text{shot})$.
	In terms of practicality, the QCNN is not easy to implement for large-scale problems requiring many qubits and a large dataset.
	Below, we present a new architecture of the QCNN that can ideally reduce the required number of shots by a factor of $\mO(1/n)$, bringing the QCNN closer to realization.

	%%%%%%%%%%%%%%%%%%%%%%%%%%%%%%%%%%%%%%%%%%%%%%%%%%%%%%%%%%%%%%%%%%%%%%%%
	%%%%%%%%%%%%%%%%%%%%%%%%%%%%%%%%%%%%%%%%%%%%%%%%%%%%%%%%%%%%%%%%%%%%%%%%
	%%%%%%%%%%%%%%%%%%%%%%%%%%%%%%%%%%%%%%%%%%%%%%%%%%%%%%%%%%%%%%%%%%%%%%%%
	\section{Split-parallelizing QCNN} \label{sec: sp-QCNN}
	
	In this section, we describe the two key components of the sp-QCNN, translationally symmetric layers and circuit splitting, and discuss the relationship between the sp-QCNN and the GQML through symmetry.
	For simplicity, this work focuses on the case where quantum data is defined on qubits aligned on a one-dimensional lattice.
	The generalization to arbitrary dimensional lattices is straightforward.

	%%%%%%%%%%%%%%%%%%%%%%%%%%%%%%%%%%%%%%%%%%%%%%%%%%%%%%%%%%%%%%%%%%%%%%%%
	\subsection{Translational symmetry}
	
	%%%%%%%%%%%%%%%%%%%%%%%%%%%%%%%%%%%%%%%%%%%%%%%%%%%%%%%%%%%%%%%%%%%%%%%%
	\begin{figure}[t]
		\centering
		\includegraphics[width=\linewidth]{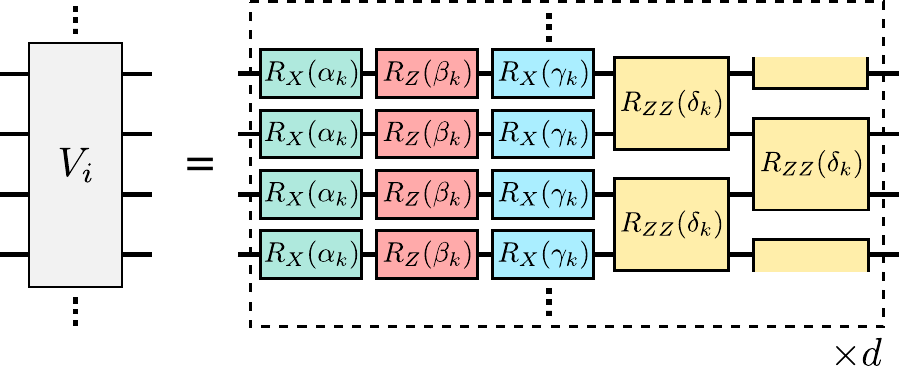}
		\caption{
			Example of translationally symmetric unitary layer.
			Single qubit rotations are applied in parallel, followed by ZZ rotations on the nearest neighboring qubits. 
			These procedures are repeated $d$ times.
			The rotation angles are translationally symmetric, and thus the number of independent parameters is $4d$.
		}
		\label{fig: Ui}
	\end{figure}
	%%%%%%%%%%%%%%%%%%%%%%%%%%%%%%%%%%%%%%%%%%%%%%%%%%%%%%%%%%%%%%%%%%%%%%%%
	
	In the sp-QCNN, we exploit data symmetry as prior knowledge to design an efficient QML model.
	The target of the sp-QCNN is translationally symmetric data, which is represented by a density matrix $\rho_i$ with the following property:
	\begin{align}
		T\rho_i T^\dag = \rho_i, \label{eq: rhoi}
	\end{align}
	where $T$ is the translation operator by one qubit (e.g., $T\ket{100\cdots}=\ket{010\cdots}$).
	The most relevant field for the application of sp-QCNN is condensed matter physics, in which translationally symmetric materials such as solids are the largest research topic~\cite{Bauer2020-uk, Arovas2022-ed, Yan2011-df}.
	In Sec.~\ref{sec: app}, we will demonstrate that the sp-QCNN can detect the quantum phases of translationally symmetric many-body states.

	To ensure the equivalence of outputs in parallel computation of sp-QCNN, we also impose translational symmetry on each of the convolutional and fully connected layers, whose unitary is denoted by $V_i$, as follows:
	\begin{align}
		T V_i T^\dag =V_i. \label{eq:VT}
	\end{align}
	As will be shown later, the translational symmetries of the data and the circuit contribute to the substantial parallel computation.
	We give an example of hardware efficient ansatz respecting the translational symmetry (Fig.~\ref{fig: Ui}):
	\begin{align}
		V_i = \prod_{k=1}^d R_{ZZ}^{\text{sym}}(\delta_k) R_{X}^{\text{sym}}(\gamma_k) R_{Z}^{\text{sym}}(\beta_k) R_{X}^{\text{sym}}(\alpha_k), \label{eq: ansatz}
	\end{align}
	where we have defined 
	\begin{align}
		R_{X}^{\text{sym}}(\theta) &= \prod_{j=1}^n e^{-i \theta X_j}, \,\,\, R_{Z}^{\text{sym}}(\theta) = \prod_{j=1}^n e^{-i \theta Z_j},\\
		R_{ZZ}^{\text{sym}}(\theta) &= \prod_{j=1}^n e^{-i \theta Z_j Z_{j+1}},
	\end{align}
	with the periodic boundary condition $Z_{j+n}=Z_{j}$.
	The rotation angles, $\alpha_k, \beta_k, \gamma_k$, and $\delta_k$, are variational parameters to be optimized and do not depend on the qubit position.
	By construction, $V_i$ is symmetric by one-qubit translation: $[V_i,T]=0$ (and therefore $[V_i,T^m]=0$ holds for any integer $m$).
	In contrast, the convolutional layer of the conventional QCNN, $V^\text{conv}_i$, is symmetric by two or more qubit translations: $[V^\text{conv}_i,T]\neq 0$ but, e.g., $[V^\text{conv}_i,T^2]= 0$ [Fig.~\ref{fig: illust}(a)].

	We note the expressivity and classical simulability of our ansatz.
	As for the expressivity, the ansatz in Eq.~\eqref{eq: ansatz} is efficient to implement in quantum hardware but is not general because it cannot express all translationally symmetric unitaries due to the extra inversion symmetry.
	In principle, the sp-QCNN allows arbitrary translationally symmetric unitaries since we do not impose any constraints on $V_i$ other than translational symmetry~\eqref{eq:VT}.
	For example, unitary operators that break inversion symmetry, such as $\exp[-i\theta \sum_j X_jY_{j+1}]$, can be implemented using Trotter decomposition~\cite{Nielsen2010-pf}.
	Then, however, the circuit tends to be deeper and therefore more difficult to implement in near-term devices.
	Also, extra symmetries of local gates can limit the circuit expressivity~\cite{Marvian2022-ir}.
	Finding a more expressive and compact ansatz is an important open issue.
	Meanwhile, it is known that strong symmetry constraints on quantum circuits can lead to classical simulability~\cite{Kerenidis2021-zl, Goh2023-lq, Cerezo2023-hz}.
	For example, permutation symmetric quantum circuits are classically simulable in many situations because the dimensions of permutation invariant subspaces are polynomial~\cite{Anschuetz2023-wn}.
	Nevertheless, since the translation symmetry is much weaker than the permutation symmetry, we suggest that the translation symmetry does not lead to classical simulability in general.
	In support of this suggestion, the dimensions of the translationally invariant subspaces are exponentially large.
	We leave further analysis on it as a future research problem. 
	%On the other hand. apart from the symmetry, a recent study claimed a conjecture that barren plateaus-free models, including QCNNs with logarithmic depth, can be classically simulable.

	%%%%%%%%%%%%%%%%%%%%%%%%%%%%%%%%%%%%%%%%%%%%%%%%%%%%%%%%%%%%%%%%%%%%%%%%
	\subsection{Circuit splitting}
	
	%%%%%%%%%%%%%%%%%%%%%%%%%%%%%%%%%%%%%%%%%%%%%%%%%%%%%%%%%%%%%%%%%%%%%%%%
	\begin{figure}[t]
		\centering
		\includegraphics[width=\linewidth]{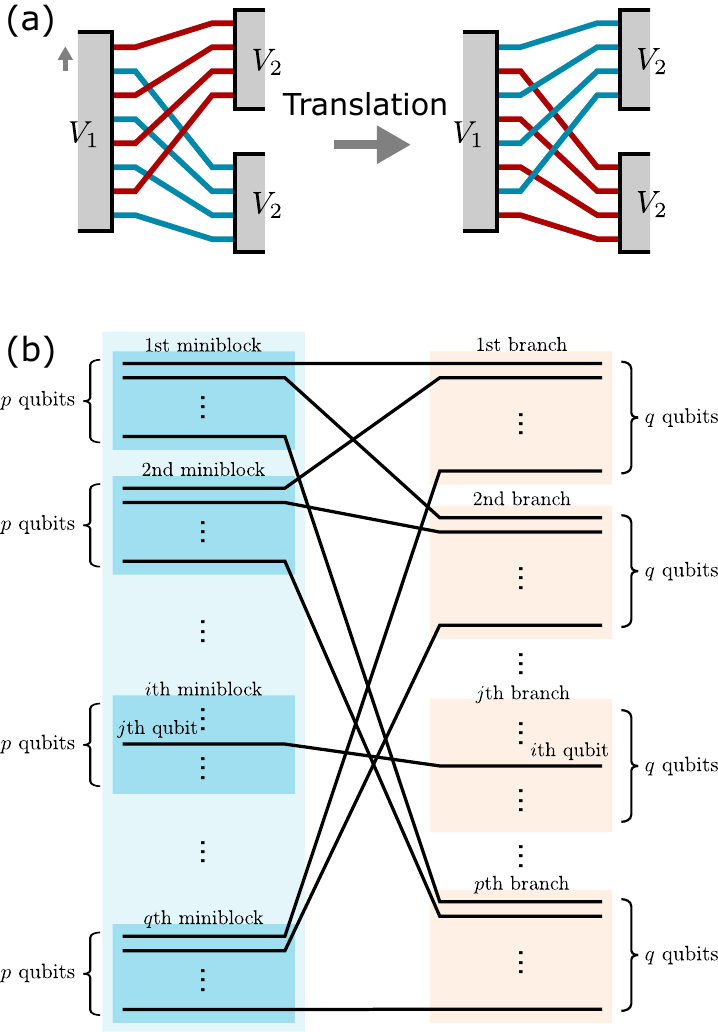}
		\caption{
			(a) An illustration of translationally symmetric circuit splitting.
			In this circuit, the entire circuit structure is invariant under the translation operation.}
		(b) A specific circuit-splitting method.
		We first divide the qubits into $q$ miniblocks consisting of $p$ qubits and split the circuit such that the $j$th qubit of the $i$th miniblock is connected to the $i$th qubit of the $j$th branch.
		\label{fig: branching_general}
	\end{figure}
	%%%%%%%%%%%%%%%%%%%%%%%%%%%%%%%%%%%%%%%%%%%%%%%%%%%%%%%%%%%%%%%%%%%%%%%%
	
	Another key component of the sp-QCNN is circuit splitting.
	In the conventional QCNN, the pooling layer discards some qubits to coarse-grain the quantum data.
	In contrast, the sp-QCNN splits the circuit at the pooling layers instead of discarding the qubits, as shown in Fig.~\ref{fig: illust}(b).
	After splitting, we perform the same operations on each branch and finally measure all the qubits in the computational basis. 
	In some types of quantum computers, such as superconducting~\cite{Song2017-pp} and ion-trap devices~\cite{Figgatt2019-sn}, unitary operations can be performed in parallel, and thus this parallel computation does not significantly increase the runtime.

	In this model, we split the circuit such that it is invariant under the translation operation.
	Figure~\ref{fig: branching_general}(a) shows an illustrative example of circuit splitting, where the translation operation only swaps the two branches (red and blue lines) but does not modify the overall circuit structure.
	Here, we give a specific circuit-splitting method for one-dimensional lattice cases.
	With $n$ as the number of qubits, we choose a prime factor of $n$, denoted by $p$, and define $q=n/p$.
	Then we introduce splitting in which $n=pq$ qubits are split into $p$ branches [Fig.~\ref{fig: branching_general}(b)].
	First, we divide the qubits into $q$ miniblocks each comprising $p$ qubits in order from top to bottom.
	Next, we split the circuit such that the $j$th qubit of the $i$th miniblock is connected to the $i$th qubit of the $j$th branch.
	By repeating this procedure on each new branch until the number of qubits becomes one, we obtain the entire sp-QCNN circuit.
	Note that this splitting procedure requires SWAP gates to rearrange the qubits in quantum hardware without all-to-all connectivity (see Appendix~\ref{sec: swap} for details).

	Due to the translational symmetry of $V_i$ and circuit splitting, the sp-QCNN substantially parallelizes the nonsplitting QCNN that consists of the same $V_i$ [Fig.~\ref{fig: illust}(b)].
	For convenience, we define $\braket{A}_\text{ns}$ and $\braket{A}$ as the expectation values of an operator $A$ in the nonsplitting and sp-QCNNs.
	In the nonsplitting QCNN, we measure one of the remaining qubits in the computational basis and consider its expectation value (i.e., $\braket{Z_1}_\text{ns}$) as the output of the QCNN.
	On the other hand, in the sp-QCNN, we measure all the qubits and regard the average of the $n$ expectation values (i.e., $\braket{Z_\text{avg}}=\sum_j \braket{Z_j}/n$) as the output.
	In the next section, we will discuss the mechanism and validity of this parallelization in more detail.

	%%%%%%%%%%%%%%%%%%%%%%%%%%%%%%%%%%%%%%%%%%%%%%%%%%%%%%%%%%%%%%%%%%%%%%%%
	\subsection{Relation with geometric quantum machine learning}

	We consider the sp-QCNN from the viewpoint of GQML or equivariant QNNs~\cite{Bronstein2021-is, Verdon2019-pi, Zheng2023-ze, Larocca2022-mj, Meyer2023-vx, Skolik2022-ge, Sauvage2024-mv, Ragone2022-va, Schatzki2022-or, Nguyen2022-go}.
	The concept of GQML has recently emerged as a potential solution to some critical QML issues associated with trainability and generalization.
	It leverages the symmetry of a problem as inductive bias and provides a problem-tailored circuit architecture.
	For example, let us consider the classical task of recognizing whether an image represents a cat.
	If an image represents a cat, then its rotated image should also represent a cat.
	In this sense, this task has rotation symmetry.
	In GQML, such symmetry is encoded in the network architecture.
	Formally, given a symmetry operation $S$ and an output function $f(\rho)$, the $S$-invariance of GQML is defined as follows:
	\begin{align}
		f(\rho) = f(S\rho S^\dag) \,\,\,\, \forall \rho. \label{eq: GQML}
	\end{align}
	In other words, the symmetry operation $S$ on the input data never changes the output of GQML.
	In GQML, the neural network is usually designed based on the equivariant circuit to satisfy this invariance. 
	In theory, GQML significantly enhances the capability of machine learning in several tasks~\cite{Zheng2023-ze, Schatzki2022-or}.

	The circuit of the sp-QCNN has the same invariant property as GQML.
	Let us consider the unitary transformation $U$ of the entire sp-QCNN.
	Due to the translational symmetry of each $V_i$ and the splitting structure, $U$ itself is translationally symmetric: 
	\begin{align}
		TUT^\dag = U. \label{eq: Usym}
	\end{align}
	This symmetry leads to the equivariant relation between input and output, $U (T\rho T^\dag) U^\dag = T (U \rho U^\dag)T^\dag$.
	That is, the translation operation applied to the input is identical to that applied to the output. 
	We also define $f(\rho)=\tr(U\rho U^\dag Z_\text{avg})$ with an observable $Z_\text{avg}=\sum_j Z_j/n$.
	Then, the equivariant relation and $[Z_\text{avg},T]=0$ result in
	\begin{align}
		f(\rho)=f(T\rho T^\dag) \,\,\,\, \forall \rho,
	\end{align}
	which is the $T$-invariance of GQML~\eqref{eq: GQML}.
	Therefore, the sp-QCNN can be seen as applying GQML to the QCNN.
	This insight suggests that the sp-QCNN can be used to enhance QML capability in tasks where the translation operation on the input data should not change the output.
	Here, let us clarify the difference between our model and the previously proposed equivariant QCNN~\cite{Nguyen2022-go}.
	In Ref.~\cite{Nguyen2022-go}, the equivariance of the pooling layer in the QCNN is achieved by randomly selecting which qubits to discard based on a given symmetry.
	The advantage of this conventional model lies in its applicability to various symmetries, while the sp-QCNN is specifically tailored to translational symmetry, imposing equivariance by splitting the circuit. 
	In terms of the number of shots required, our model should be more efficient than the conventional equivariant QCNN by maximizing the utilization of qubit resources.

	Our work also provides a new direction for exploiting data symmetry to improve the potential of QML.
	A critical difference between our problem setting and common ones in GQML is that the input data itself is symmetric in our problem [Eq.~\eqref{eq: rhoi}], but not in GQML (e.g., the cat image is not rotation invariant).
	Therefore, each approach brings different benefits.
	Although the usual GQML improves trainability and generalization, our method reduces measurement costs through substantial parallelization.
	Thus, the sp-QCNN is particularly advantageous for near-term quantum devices where computational resources are limited.

	%%%%%%%%%%%%%%%%%%%%%%%%%%%%%%%%%%%%%%%%%%%%%%%%%%%%%%%%%%%%%%%%%%%%%%%%
	\begin{figure*}[t]
		\centering
		\includegraphics[width=\linewidth]{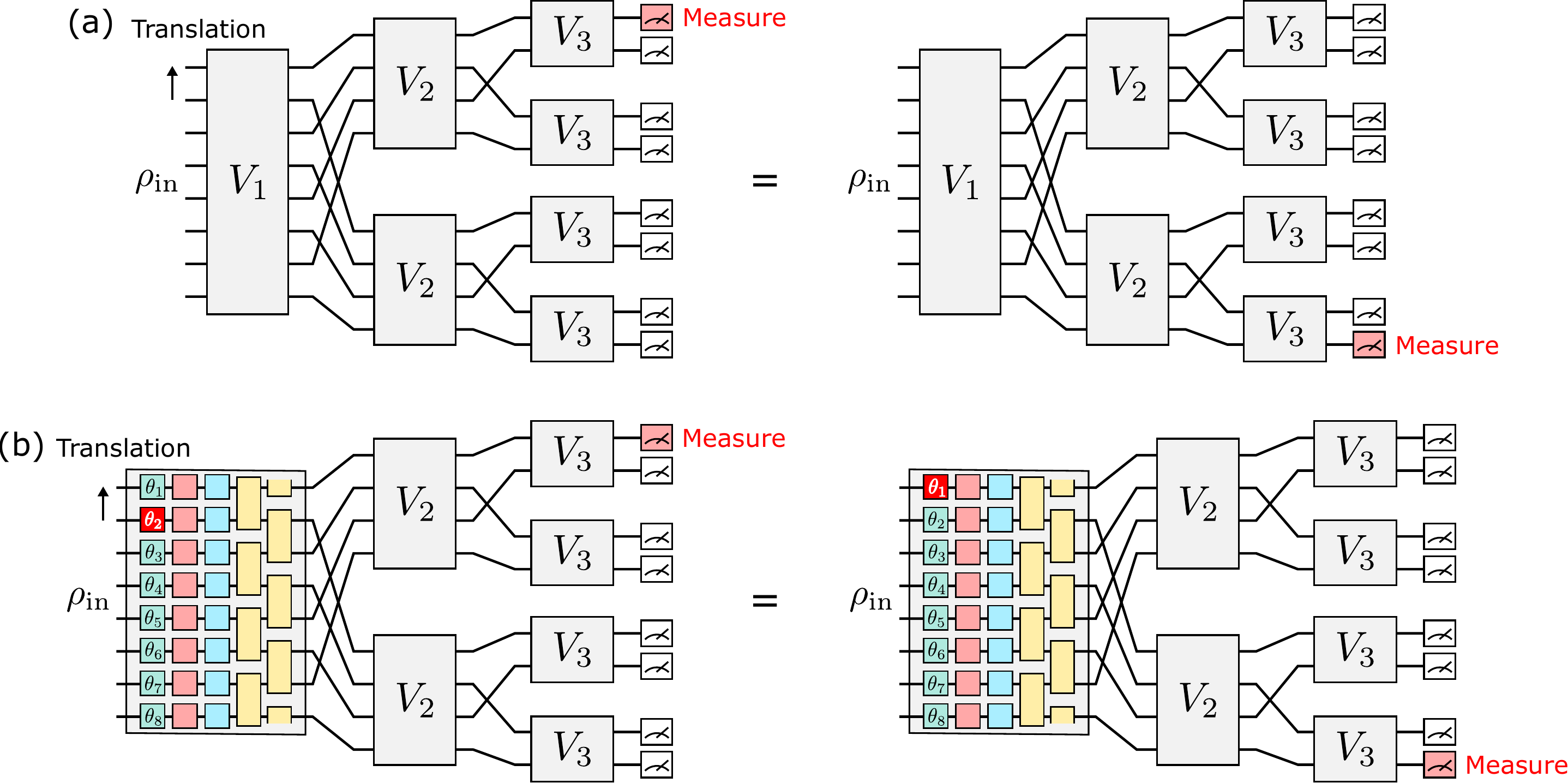}
		\caption{
			Mechanism of parallelization in the sp-QCNN.
			(a) In the sp-QCNN, the expectation value of a local observable is equivalent for all the qubits.
			This can be proved by virtually translating the entire circuit. 
			The translation does not change the input state and quantum circuit due to their translational symmetry but shifts the position of the measured qubit, showing the equivalence of expectation values at different qubits.
			(b) The gradient measurement can be parallelized in the sp-QCNN.
			In accordance with the chain rule, the gradient is the sum of several derivatives, $\partial \braket{Z_1}/\partial \theta = \sum_j \partial \braket{Z_1}/\partial \theta_j$.
			For example, we suppose that the parameter $\theta$ is in the first convolutional layer as shown in the figure (the red boxes denote $\partial/\partial \theta_2$ and $\partial/\partial \theta_1$).
			Then translating the circuit proves $\partial \braket{Z_1}/\partial \theta_j = \partial \braket{Z_{j-2}}/\partial \theta_1$ and thus $\partial \braket{Z_1}/\partial \theta = \sum_j \partial \braket{Z_j}/\partial \theta_1$, which can be computed with only two circuits by measuring all the qubits.
		}
		\label{fig: meachnism}
	\end{figure*}
	%%%%%%%%%%%%%%%%%%%%%%%%%%%%%%%%%%%%%%%%%%%%%%%%%%%%%%%%%%%%%%%%%%%%%%%%

	%%%%%%%%%%%%%%%%%%%%%%%%%%%%%%%%%%%%%%%%%%%%%%%%%%%%%%%%%%%%%%%%%%%%%%%%
	%%%%%%%%%%%%%%%%%%%%%%%%%%%%%%%%%%%%%%%%%%%%%%%%%%%%%%%%%%%%%%%%%%%%%%%%
	%%%%%%%%%%%%%%%%%%%%%%%%%%%%%%%%%%%%%%%%%%%%%%%%%%%%%%%%%%%%%%%%%%%%%%%%
	\section{Measurement efficiency in sp-QCNN} \label{sec: mechanism}
	
	In this section, we describe the parallelization mechanism in the sp-QCNN and show that it can improve the measurement efficiency of local observables and their gradients.
	We also analytically prove that the improvement factor is $\mO(n)$ for a random input state.

	\subsection{Measurement efficiency of local observable}

	First, we show that the translational symmetry of $V_i$ and circuit splitting allow for parallel computation and improve the measurement efficiency of local observables.
	A key property of the sp-QCNN is the equivalence of expectation values for all the qubits.
	We recall that the unitary transformation $U$ of the entire sp-QCNN is translationally symmetric [Eq.~\eqref{eq: Usym}].
	This symmetry leads to 
	\begin{align}
		\braket{Z_1} 
		&= \tr\left(U\rho U^\dag Z_1\right) \notag \\
		&= \tr\left(U (T^\dag)^{j-1} \rho T^{j-1} U^\dag Z_1\right) \notag \\
		&= \tr\left(U\rho U^\dag Z_{j}\right) \notag \\
		&= \braket{Z_j}, \label{eq: equiv}
	\end{align}
	where $\rho$ is an input state satisfying Eq.~\eqref{eq: rhoi}, and we have used $\rho=(T^\dag)^{j-1} \rho T^{j-1}$ and $T^{j-1} Z_1 (T^\dag)^{j-1} = Z_j$.
	This equation indicates the equivalence of the expectation values for all the qubits, i.e., $\braket{Z_i} =\braket{Z_j}$ for any $i$ and $j$.
	This argument can be applied to other single-qubit Pauli operators, leading to $\braket{X_i} =\braket{X_j}$ and $\braket{Y_i} =\braket{Y_j}$.
	Figure~\ref{fig: meachnism}(a) graphically illustrates this equivalence, which can also be proved by translating the circuit.

	This equivalence tells us that the sp-QCNN substantially parallelizes the nonsplitting QCNN that consists of the same $V_i$, as shown in Fig.~\ref{fig: illust}(b).
	As mentioned above, we regard the average of the expectation values for all the qubits, $\braket{Z_\text{avg}}=\sum_j \braket{Z_j}/n$, as the output in the sp-QCNN.
	Meanwhile, we consider the expectation value for only one qubit, $\braket{Z_1}_\text{ns}$, as the output in the nonsplitting QCNN.
	Given the equivalence in Eq.~\eqref{eq: equiv}, the nonsplitting and sp-QCNNs produce the same results if statistical errors are absent:
	\begin{align}
		\braket{Z_1}_\text{ns} = \braket{Z_\text{avg}}. \label{eq: z1zavg}
	\end{align}
	Here we have used $\braket{Z_1}_\text{ns} = \braket{Z_1}$, which can be proved by noticing that the nonsplitting QCNN is a part of the sp-QCNN.
	In the sp-QCNN, we estimate the output from $T$ measurement shots as follows:
	\begin{align}
		\braket{Z_\text{avg}}_\text{est} = \frac{1}{T} \sum_{\ell=1}^T z_\text{avg}^{(\ell)}=\frac{1}{nT} \sum_{\ell=1}^T \sum_{j=1}^n z_j^{(\ell)}. \label{eq: Z_est}
	\end{align}
	Here, $z_j^{(\ell)}=\pm1$ is the $\ell$th measurement outcome at the $j$th qubit, and we have defined the average of the $\ell$th measurement outcomes as $z_\text{avg}^{(\ell)}=\sum_j z_j^{(\ell)}/n$.
	The value of $z_\text{avg}^{(\ell)}$ can be $a/n$ ($a\in\{-n,-n+2,\cdots,n\}$), corresponding to the measurement outcome of $Z_\text{avg}$.
	We note that the number of outcomes in the sp-QCNN is $n$ times greater than that in the nonsplitting QCNN where the output is estimated as $\sum_{\ell=1}^T z_1^{(\ell)}/T$.
	Therefore, the sp-QCNN can reduce the number of shots required to achieve a certain estimation accuracy.
	Since this argument only relies on the symmetry property of data, the sp-QCNN is general and can be applied to broad tasks with translationally symmetric data.

	It is worth noting that the sp-QCNN does not necessarily improve the measurement efficiency by a factor of $\mO(n)$.
	This is because, in each shot, $n$ measurement outcomes are correlated to each other via quantum entanglement.
	For example, if the output state is the GHZ state $\ket{\psi}=(\ket{000\cdots}+\ket{111\cdots})/\sqrt{2}$, then the sp-QCNN does not improve the measurement efficiency at all because the $n$ outcomes are completely correlated and can only provide one bit of information.
	In contrast, if the output state is the W state $\ket{\psi}=(\ket{100\cdots00}+\ket{010\cdots00}+\cdots+\ket{000\cdots01})/\sqrt{n}$, then the exact expectation value can be obtained with only one shot by measuring all the qubits in the sp-QCNN, whereas many measurements are required in the nonsplitting QCNN.
	Therefore, how well the sp-QCNN improves the measurement efficiency depends on the details of the problem, such as input data and circuit parameters.
	Later, we will analytically prove that the sp-QCNN can improve the measurement efficiency by a factor of $\mO(n)$ for a typical random input state.

\begin{figure}[t]
		\centering
		\includegraphics[width=\linewidth]{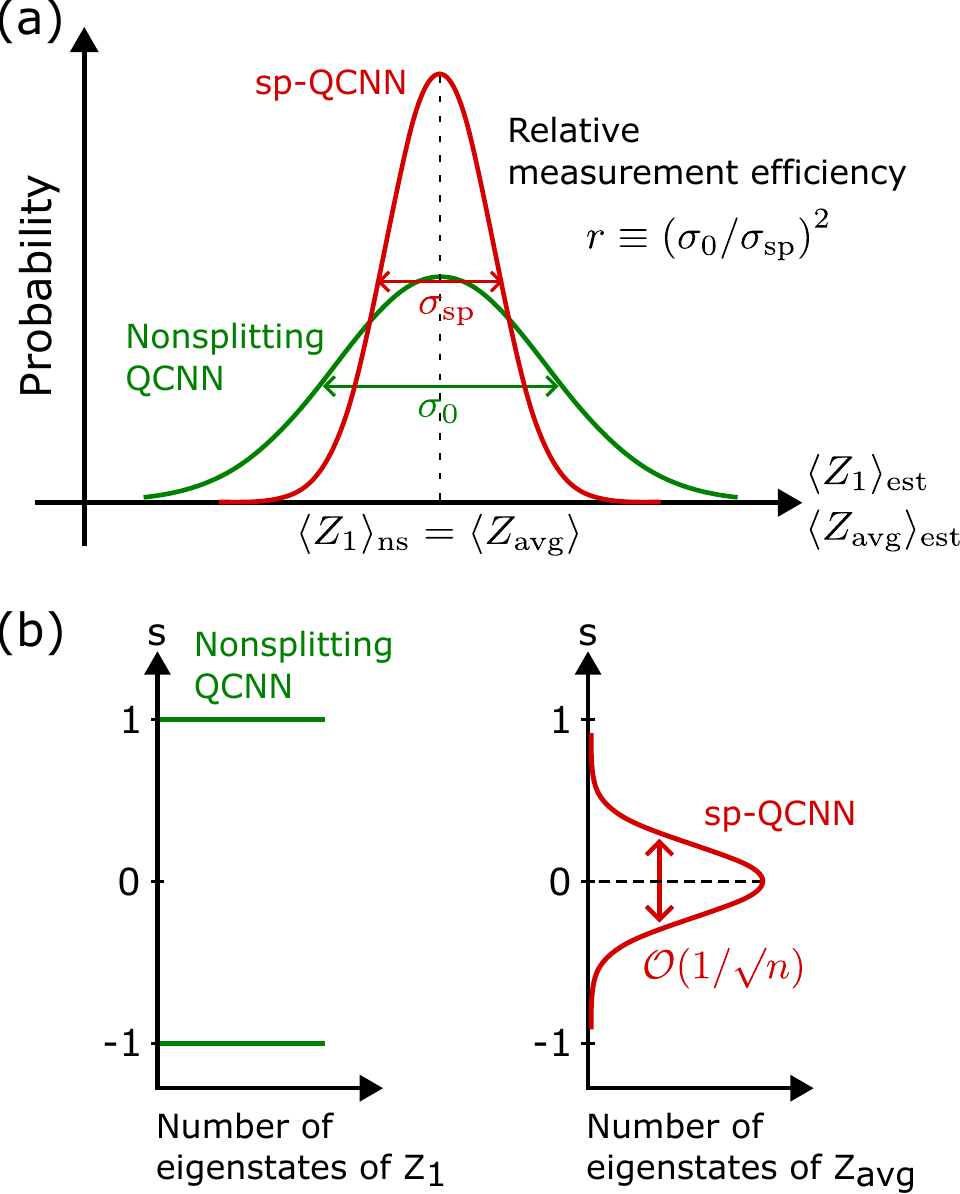}
		\caption{
			(a) Quantification of measurement efficiency.
			In actual experiments, statistical errors arise in estimating the expectation value of an observable.
			This figure shows the probability distribution of the estimated expectation value.
			Here, we define the relative measurement efficiency $r$ as the ratio of the variances in the sp-QCNN and the nonsplitting QCNN.
			(b) Number of eigenstates of $Z_1$ and $Z_\text{avg}$ with an eigenvalue $s$.
			While the possible measurement outcome is $\pm1$ in the nonsplitting QCNN (left panel), it is widely distributed in the range of $-1$ to 1 with a width of $\mO(1/\sqrt{n})$ in the sp-QCNN (right panel).
		}
		\label{fig: variance}
	\end{figure}

	The advantage of the sp-QCNN is illustrated in Fig.~\ref{fig: variance}(a).
	In actual experiments, we cannot obtain the exact expectation value because of statistical errors.
	Therefore it is usually estimated from the mean value of a finite number of measurement outcomes.
	In the nonsplitting QCNN, the estimated value is generally drawn from the Gaussian distribution with a variance of $\mO(1/N_\text{shot})$ in accordance with the central limit theorem.
	In the sp-QCNN, we obtain $n$ measurement outcomes at once and thus expect that the variance scales as $\mO(1/nN_\text{shot})$, indicating the $\mO(n)$ times improvement of measurement efficiency.
	We note that the sp-QCNN can improve the measurement efficiency of the conventional QCNN, not necessarily other QNNs.
	Our model further enhances the feasibility of QCNNs, bringing its practical quantum advantages closer to realization.

	To quantify the effectiveness of the sp-QCNN, we introduce the relative measurement efficiency:
	\begin{align}
		r \equiv \left( \frac{\sigma_0}{\sigma_\text{sp}} \right)^2. \label{eq: r}
	\end{align}
	Here $\sigma_0$ and $\sigma_\text{sp}$ are the standard deviations (i.e., square root of variance) of the Gaussians followed by estimated expectation values in the nonsplitting and sp-QCNNs with the same number of shots.
	This quantity means that the shot number required to achieve a certain estimation accuracy using the sp-QCNN is $1/r$-times fewer than that using the nonsplitting QCNN.
	In the next section, we will demonstrate the efficiency of the sp-QCNN for a concrete task using this quantity.

	%%%%%%%%%%%%%%%%%%%%%%%%%%%%%%%%%%%%%%%%%%%%%%%%%%%%%%%%%%%%%%%%%%%%%%%%
	%%%%%%%%%%%%%%%%%%%%%%%%%%%%%%%%%%%%%%%%%%%%%%%%%%%%%%%%%%%%%%%%%%%%%%%%

	%%%%%%%%%%%%%%%%%%%%%%%%%%%%%%%%%%%%%%%%%%%%%%%%%%%%%%%%%%%%%%%%%%%%%%%%
	\subsection{Measurement efficiency of gradient} \label{sec: grad_eff}
	In general, the most costly part of machine learning is the optimization of neural networks using a training dataset, in which the loss function is minimized by tuning the network parameters.
	In classical machine learning, gradient-based methods are often used for optimization and work well for large-scale problems.
	Even in QML, gradient-based optimizers are important and powerful tools.
	However, many measurements are necessary to estimate the gradient in quantum computing~\cite{Abbas2023-oo}.
	Our architecture makes such gradient measurements efficient.

	We first describe a conventional way of measuring the gradient of $\braket{Z_1}$.
	In many QCNNs, including the sp-QCNN, multiple quantum gates share a single variational parameter $\theta$.
	Here, let $m_\theta$ be the number of gates sharing $\theta$ in a branch.
	To calculate the gradient by $\theta$, we suppose that the $m_\theta$ gates have different variational parameters from each other, $\theta_j$ ($j=1,\cdots,m_\theta$).
	Thereby, we calculate the gradient with the chain rule as
	$\partial \braket{Z_1}/\partial \theta = \sum_j (\partial \theta_j/\partial \theta)(\partial\braket{Z_1}/\partial \theta_j) = \sum_j \partial\braket{Z_1}/\partial \theta_j$,
	where we have used $\partial \theta_j/\partial \theta=1$. 
	When each gate is parametrized as $e^{-i\theta_j P}$ ($P$ is a Pauli operator), $\partial\braket{Z_1}/\partial \theta_j$ can be measured using the parameter-shift rule, $\partial\braket{Z_1}/\partial \theta_j = \braket{Z_1}_{\theta_j=\theta+\pi/4} - \braket{Z_1}_{\theta_j=\theta-\pi/4} $~\cite{Mitarai2018-ap, Schuld2019-rr}.
	Thus, in the sp-QCNN, we can compute the gradient as follows:
	\begin{align}
		\frac{\partial \braket{Z_1}}{\partial \theta} 
		&= \sum_{j=1}^{m_\theta}  \tr\left( \tilde{U}_{j_+} \rho \tilde{U}_{j_+}^\dag Z_1 \right) - \tr\left( \tilde{U}_{j_-} \rho \tilde{U}_{j_-}^\dag Z_1 \right), \label{eq: bpqcnn_grad0}
	\end{align}
	where $\tilde{U}_{j_\pm}$ is the unitary transformation of the sp-QCNN in which $\theta_j=\theta$ is replaced with $\theta_j=\theta\pm\pi/4$.
	This formula has $2m_\theta$ terms, each of which is usually measured in a different circuit.

	The circuit splitting and translational symmetry in the sp-QCNN allow us to compute $\partial\braket{Z_1}/\partial \theta_j$ in parallel, improving the gradient measurement efficiency.
	For simplicity, we suppose that each $V_i$ has the form in Eq.~\eqref{eq: ansatz} and that $\theta$ is in the first convolutional layer [Fig.~\ref{fig: meachnism}(b)].
	By translating the entire circuit, we can rewrite each term in Eq.~\eqref{eq: bpqcnn_grad0} as
	\begin{align}
		\tr\left( \tilde{U}_{j_\pm} \rho \tilde{U}_{j_\pm}^\dag Z_1 \right) 
		&= \tr\left( \tilde{U}_{j_\pm} T^{j-1}\rho (T^\dag)^{j-1} \tilde{U}_{j_\pm}^\dag Z_1 \right) \notag \\
		&= \tr\left( \tilde{U}_{1_\pm} \rho \tilde{U}_{1_\pm}^\dag Z_{2-j} \right).
	\end{align}
	Here we have used $\rho=T^{j-1} \rho (T^\dag)^{j-1}$, $(T^\dag)^{j-1} Z_1 T^{j-1} = Z_{2-j}$, and $(T^\dag)^{j-1} \tilde{U}_{j_\pm} T^{j-1}=\tilde{U}_{1_\pm}$.
	This relation tells us that the derivative of $Z_1$ by $\theta_j$ is identical to that of $Z_{2-j}$ by $\theta_1$, as illustrated in Fig.~\ref{fig: meachnism}(b).
	Thereby, Eq.~\eqref{eq: bpqcnn_grad0} is reduced to
	\begin{align}
		\hspace{-0.1cm}\frac{\partial \braket{Z_1}}{\partial \theta} 
		= \sum_{j=1}^{m_\theta} \tr\left( \tilde{U}_{1_+} \rho \tilde{U}_{1_+}^\dag Z_j \right) - \tr\left( \tilde{U}_{1_-} \rho \tilde{U}_{1_-}^\dag Z_j \right),
	\end{align}
	where we have replaced $Z_{j-2}$ with $Z_{j}$ in the summation.
	According to this equation, we can obtain the gradient $\partial\braket{Z_1}/\partial \theta$ with just two circuits $\tilde{U}_{1\pm}$ by measuring all the qubits, instead of using $2m_\theta$ circuits that are conventionally necessary.

	By generalizing this argument and using the equivalence $\braket{Z_\text{avg}}=\braket{Z_1}$, we estimate the gradient of the output as follows:
	\begin{align}
		\left( \frac{\partial \braket{Z_\text{avg}}}{\partial \theta}\right)_\text{est} 
		&= \frac{m_\theta}{nT} \sum_{\ell=1}^T \sum_{j=1}^n \left[ z_{j+}^{(\ell)} - z_{j-}^{(\ell)} \right],
	\end{align}
	where $z_{j\pm}^{(\ell)}$ is the $j$th qubit measurement outcome of the $\ell$th shot in the parameter-shifted circuit with $\theta_1=\theta\pm\pi/4$.
	In our ansatz [Eq.~\eqref{eq: ansatz}], the factor $m_\theta/n$ appears when $\theta$ is in the second or later layer.
	We emphasize that the sp-QCNN enables us to execute $n$ parallel computations even for the gradient estimation, thus accelerating the gradient-based training.
	Similar to the previous case, the relative measurement efficiency $r$ for the gradient depends on the details of the problem due to the entangled property of the output state.

	%%%%%%%%%%%%%%%%%%%%%%%%%%%%%%%%%%%%%%%%%%%%%%%%%%%%%%%%%%%%%%%%%%%%%%%%
	\subsection{Measurement efficiency for random state} \label{sec: random}

	How well the sp-QCNN improves the measurement efficiency depends on the details of the problem.
	Here, we analytically prove that the efficiency is improved by a factor of $\mO(n)$ for a typical state randomly chosen from the $T$-invariant Hilbert subspace in the limit of $n\to\infty$.

	Let us begin by considering the nonsplitting QCNN, where we measure $Z_1$ and obtain an outcome $s=\pm1$ for every measurement.
	In the limit of $n\to\infty$, the probability of obtaining an outcome $\pm1$ is almost $1/2$ for a typical random state because the statistical fluctuations by randomness are negligible due to the exponentially large Hilbert space [this probability distribution is depicted in the left panel of Fig.~\ref{fig: variance}(b)]. 
	Given its Bernoulli distribution, the estimation accuracy of the expectation value is 
	\begin{align}
		\sigma_0 \sim \mO \left( \frac{1}{\sqrt{N_\text{shot}}} \right),
	\end{align}
	where $N_\text{shot}$ is the number of shots.

	In the sp-QCNN, we measure all the qubits in the computational basis and regard the mean of the $n$ measurement outcomes as the output of the QCNN [Eq.~\eqref{eq: Z_est}].
	In other words, we measure $Z_\text{avg} = \sum_j Z_j/n$ rather than $Z_1$ and obtain one of the eigenvalues $s$ ($=\pm1,\pm(n-2)/n,\cdots$) as an outcome.
	Also, given that the full unitary transformation $U$ is translationally symmetric, the output state of the sp-QCNN has the same symmetry.
	The right panel of Fig.~\ref{fig: variance}(b) shows the number of eigenstates of $Z_\text{avg}$ with an eigenvalue $s$, $D_n(s)$, on the $T$-invariant Hilbert subspace.
	In the limit of $n\to\infty$, $D_n(s)$ approaches the following asymptotic form (see Appendix~\ref{sec: DOS} for derivation): 
	\begin{align}
		D_n(s) \sim \frac{C_n}{(1+s^2)^{n/2}}, \label{eq: Dn}
	\end{align}
	where $C_n$ is a constant independent of $s$.
	The width of $D_n(s)$ in $s$ is $\mO(1/\sqrt{n})$, which finally gives rise to a small estimation error.

	Here, we assume that when measuring $Z_\text{avg}$ for a typical state randomly chosen from the $T$-invariant subspace, the probability of obtaining an outcome $s$ is proportional to $D_n(s)$.
	This assumption would be justified in the limit of $n\to\infty$, where $D_n(s)$ is sufficiently large, and the statistical fluctuations are insignificant.
	Considering that the width of $D_n(s)$ is $\mO(1/\sqrt{n})$, we can estimate the expectation value from $N_\text{shot}$ experiments with an accuracy
	\begin{align}
		\sigma_\text{sp} \sim \mO \left( \frac{1}{\sqrt{nN_\text{shot}}} \right).
	\end{align}
	From the quantification in Eq.~\eqref{eq: r}, the relative measurement efficiency of the sp-QCNN is 
	\begin{align}
		r = \left( \frac{\sigma_0}{\sigma_\text{sp}} \right)^2 \sim \mO(n). \label{eq: scaling}
	\end{align}
	This result indicates the $\mO(1/n)$ times reduction in the number of experiments required to achieve a certain accuracy.

	The scaling argument in Eq.~\eqref{eq: scaling} is valid in situations where the output state is a random quantum state. 
	Therefore, it may arise in the early stage of the learning process when the parameters of the QCNN are randomly initialized and the output state is approximately random.
	In Sec.~\ref{sec: app}, we will show that the sp-QCNN exhibits $\mO(n)$ scaling for a concrete task in the early stage of learning and, remarkably, even in the final stage.

	%%%%%%%%%%%%%%%%%%%%%%%%%%%%%%%%%%%%%%%%%%%%%%%%%%%%%%%%%%%%%%%%%%%%%%%%
	%%%%%%%%%%%%%%%%%%%%%%%%%%%%%%%%%%%%%%%%%%%%%%%%%%%%%%%%%%%%%%%%%%%%%%%%
	%%%%%%%%%%%%%%%%%%%%%%%%%%%%%%%%%%%%%%%%%%%%%%%%%%%%%%%%%%%%%%%%%%%%%%%%
	\section{Application to quantum phase recognition} \label{sec: app}
	
	In this section, we apply the sp-QCNN to a quantum phase recognition task investigated in Ref.~\cite{Cong2019-ov} and verify its effectiveness.
	For the remainder of this paper, we simulate the quantum circuit with Qulacs, an open-source quantum circuit simulator~\cite{Suzuki2021-uh}.

	%%%%%%%%%%%%%%%%%%%%%%%%%%%%%%%%%%%%%%%%%%%%%%%%%%%%%%%%%%%%%%%%%%%%%%%%
	\subsection{Formulation of problem}
	
	Let us consider a one-dimensional cluster Ising model with the periodic boundary condition, whose Hamiltonian is given by
	\begin{align}
		H = -\sum_{j=1}^n Z_j X_{j+1} Z_{j+2} - h_1\sum_{j=1}^n X_j - h_2\sum_{j=1}^n X_j X_{j+1}, \label{eq: Ham}
	\end{align}
	where $n$ is the number of qubits, and $X_j, Y_j$, and $Z_j$ are the Pauli operators at the $j$th qubit.
	This Hamiltonian exhibits SPT~\cite{Gu2009-zd, Pollmann2010-ed, Chen2011-yh, Pollmann2012-lg}, paramagnetic (PM), and antiferromagnetic (AFM) phases on the $h_1$-$h_2$ plane.
	The SPT phase is protected by $\mathbb{Z}_2\times\mathbb{Z}_2$ symmetry characterized by $X_\text{even(odd)}=\prod_{j\in \text{even(odd)}}X_j$.
	The ground state of $H$, an input state in our task, is translationally symmetric because of $THT^\dag = H$.

	Our task is to recognize the SPT phase using the sp-QCNN.
	Quantum phase recognition is one of the main applications of the QCNN, and many studies have been conducted with the aim of practical quantum advantages~\cite{Herrmann2022-bs, Liu2022-uu, Monaco2023-xr}.
	In this task, the sp-QCNN can be applied because the input data (i.e., the ground state of $H$) is translationally symmetric.
	For training data, we use $20$ ground states of $H$ evenly located on the line of $h_2=0$ from $h_1=0.05$ to $1.95$.
	Using the Jordan--Wigner transformation~\cite{Jordan1928-ic}, we can analytically obtain the exact ground state for $h_2=0$, which transits from the SPT to PM phases at $h_1=1$.  
	To evaluate the generalization of our method, we test the trained model with 28 data samples, most of which are not included in the training dataset. 
	These samples correspond to ground states at $(h_1,h_2)$ with $h_1\in\{0.35,0.65,0.95,1.25\}$ and $h_2\in\{0,\pm0.5,\pm1,\pm1.5\}$.
	The exact phase diagram on the $h_1$-$h_2$ plane is computed with the density matrix renormalization group (DMRG)~\cite{White1992-sw, White1993-qw, Schollwock2011-zv, Fishman2022-lh}.
	In this work, we prepare the input data with exact diagonalization for simplicity.
	Yet, in actual experiments, other preparation methods must be applied, such as variational quantum eigensolver on a quantum computer and analog-digital transduction from a quantum experiment.

	To train our model and evaluate its generalization, we consider the following error as the loss function:
	\begin{align}
		\mathcal{L} = \frac{1}{2M} \sum_{i=1}^M \left( \braket{\phi_i | U^\dag Z_\text{avg} U | \phi_i} - y_i \right)^2,
	\end{align}
	where $\ket{\phi_i}$ and $y_i$ are the training/test data and its corresponding label, and $U$ is the total unitary of the circuit ($M$ is the number of training/test data).
	Here, we set $y_i$ as 1 if $\ket{\phi_i}$ belongs to the SPT phase and 0 if it does not.
	We optimize the loss function using the stochastic gradient descent (SGD) method~\cite{Robbins1951-ql}.
	In SGD, we update the parameters as $\vec{\theta}^{(t+1)} = \vec{\theta}^{(t)} - \eta^{(t)}\nabla \mathcal{L}$, where $\vec{\theta}^{(t)}$ is the parameter vector at optimization step $t$, and $\nabla \mathcal{L}$ is calculated from only one of the training data at each step.
	We also decrease the learning rate as $\eta^{(t)}=\eta_0/t$ to stabilize the training and set $\eta_0=200$.
	Besides, to investigate the statistical properties of the sp-QCNN, we simulate the same circuits with $N_p$ different random initial parameter sets. 
	We set $N_p=50$ in Sec.~\ref{sec: performance} and $N_p=200$ in Sec.~\ref{sec: eff}.

	%%%%%%%%%%%%%%%%%%%%%%%%%%%%%%%%%%%%%%%%%%%%%%%%%%%%%%%%%%%%%%%%%%%%%%%%
	\subsection{Performance with limited measurement resources}\label{sec: performance}

	%%%%%%%%%%%%%%%%%%%%%%%%%%%%%%%%%%%%%%%%%%%%%%%%%%%%%%%%%%%%%%%%%%%%%%%%
	\begin{figure*}[t]
		\centering
		\includegraphics[width=\linewidth]{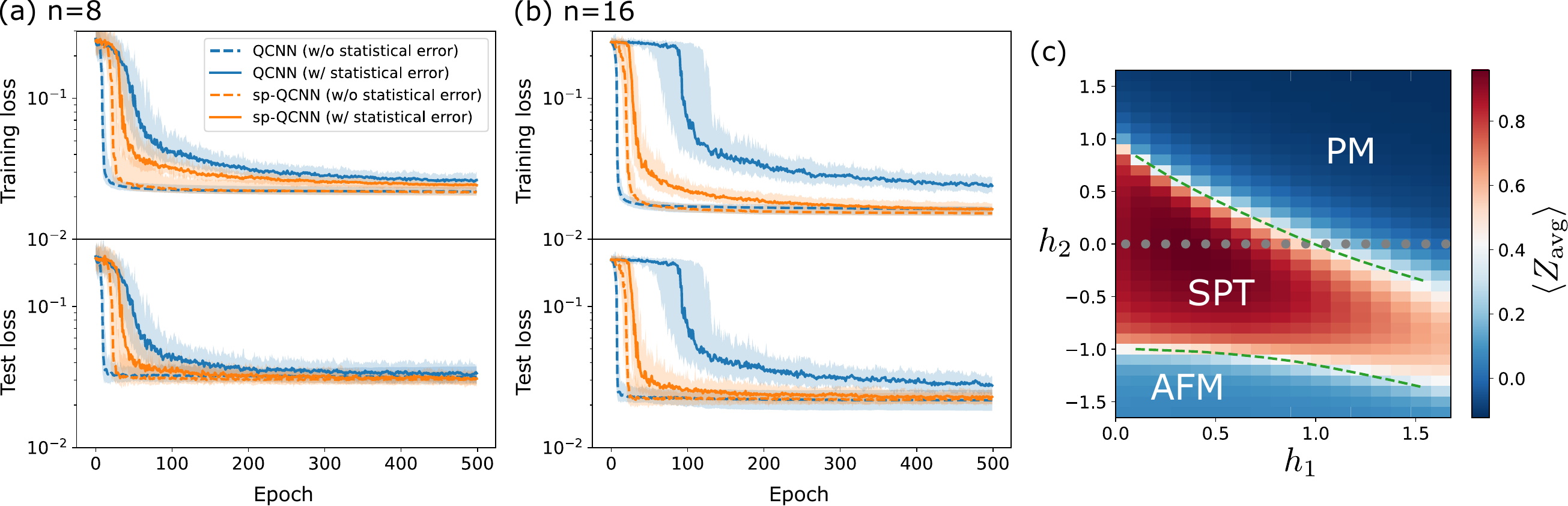}
		\caption{
			(a), (b) Changes in training (top) and test (bottom) loss functions for (a) $n=8$ and (b) $n=16$.
			The orange (blue) solid and dashed lines denote the loss functions with and without statistical errors in the sp-QCNN (conventional QCNN) respectively. 
			The shaded areas are the 10th–90th percentiles of the loss function for 50 sets of random initial parameters at each epoch. 
			We match the number of shots per parameter to obtain the gradient in both QCNNs.
			(c) Phase diagram predicted by the trained sp-QCNN with statistical errors for $n=16$ qubits.
			The color denotes the average magnitude of $\braket{Z_\text{avg}}$ for 50 sets of initial parameters.
			The gray dots and dashed green lines denote our training data and phase boundaries computed by DMRG, respectively.
			In these simulations, we set the depth of each layer as $d=5$.
		}
		\label{fig: fixedshots}
	\end{figure*}
	%%%%%%%%%%%%%%%%%%%%%%%%%%%%%%%%%%%%%%%%%%%%%%%%%%%%%%%%%%%%%%%%%%%%%%%%

	We investigate how the high measurement efficiency of sp-QCNN enhances the machine learning performance in training with limited measurement resources.
	In such situations, statistical errors in estimating the gradient of the loss function would disturb the training process,  reducing accuracy within a limited computational resource.
	Here, we show that the sp-QCNN can suppress the statistical errors, stabilizing and speeding up the learning process.

	As the sp-QCNN circuit, we use the ansatz in Eq.~\eqref{eq: ansatz} with $d=5$, where the total number of parameters is $60$ for $n=8$ and $80$ for $n=16$.
	We also compare the sp-QCNN with the conventional QCNN depicted in Fig.~\ref{fig: illust}(a), where the convolutional and fully-connected layers consist of two-qubit unitary gates parametrized as $\prod_{j=1}^{15} e^{-i \theta_j P_j}$ ($P_j=IX,IY,\cdots,ZZ$).
	Since the gates acting in parallel share the same parameters, the number of independent parameters in the conventional QCNN is $75$ for $n=8$ and $105$ for $n=16$.
	When measuring the gradient in the simulation, we match the shot number per parameter in the conventional and sp-QCNNs for each layer.
	Here, we use $2m_\theta N_\text{shot}$ shots for parameter $\theta$ in the sp-QCNN and set $N_\text{shot}=5$.

	We first show that our sp-QCNN has comparable classification performance to the conventional QCNN.
	Figures~\ref{fig: fixedshots}(a) and (b) illustrate the changes in the training and test loss functions with and without the statistical errors of gradients for the conventional and sp-QCNNs.
	Both with and without statistical errors, the final values of the training and test losses in the sp-QCNN (orange lines) are equal to or better than those of the conventional one (blue lines).
	This result indicates that our model exhibits sufficient expressivity and generalization for this problem compared to the conventional QCNN.
	Furthermore, our model can accurately predict the unknown phase diagram.  
	Figure~\ref{fig: fixedshots}(c) presents the phase diagram predicted by the trained sp-QCNN with statistical errors for $n = 16$ qubits. 
	It closely aligns with the phase boundary computed by DMRG (shown as dashed lines). 
	Notably, the sp-QCNN can identify the SPT-AFM phase transition, even though this phase transition was not part of the training dataset.

	The sp-QCNN effectively suppresses statistical errors and accelerates the learning process while maintaining comparable classification performance to the conventional QCNN.
	In Figs.~\ref{fig: fixedshots}(a) and (b), the loss functions converge rapidly for both QCNNs in the absence of statistical errors (dashed lines).
	However, in the presence of statistical errors (solid lines), the loss convergence becomes significantly slower in the conventional QCNN, whereas it remains relatively modest in the sp-QCNN.
	This fast convergence in the sp-QCNN stems from its high measurement efficiency.
	While significant statistical errors disturb the rapid and stable optimization in the conventional QCNN, the high measurement efficiency in the sp-QCNN suppresses the statistical errors, stabilizing and accelerating the optimization.
	As shown in Figs.~\ref{fig: fixedshots}(a) and (b), this improvement is more prominent for $n = 16$ compared to $n = 8$, due to the $\mO(n)$ improvement in the measurement efficiency.
	The fast convergence of training is highly effective for near-term quantum devices where a long optimization run is impractical due to limited computational resources.

	%%%%%%%%%%%%%%%%%%%%%%%%%%%%%%%%%%%%%%%%%%%%%%%%%%%%%%%%%%%%%%%%%%%%%%%%
	\begin{figure*}[t]
		\centering
		\includegraphics[width=\linewidth]{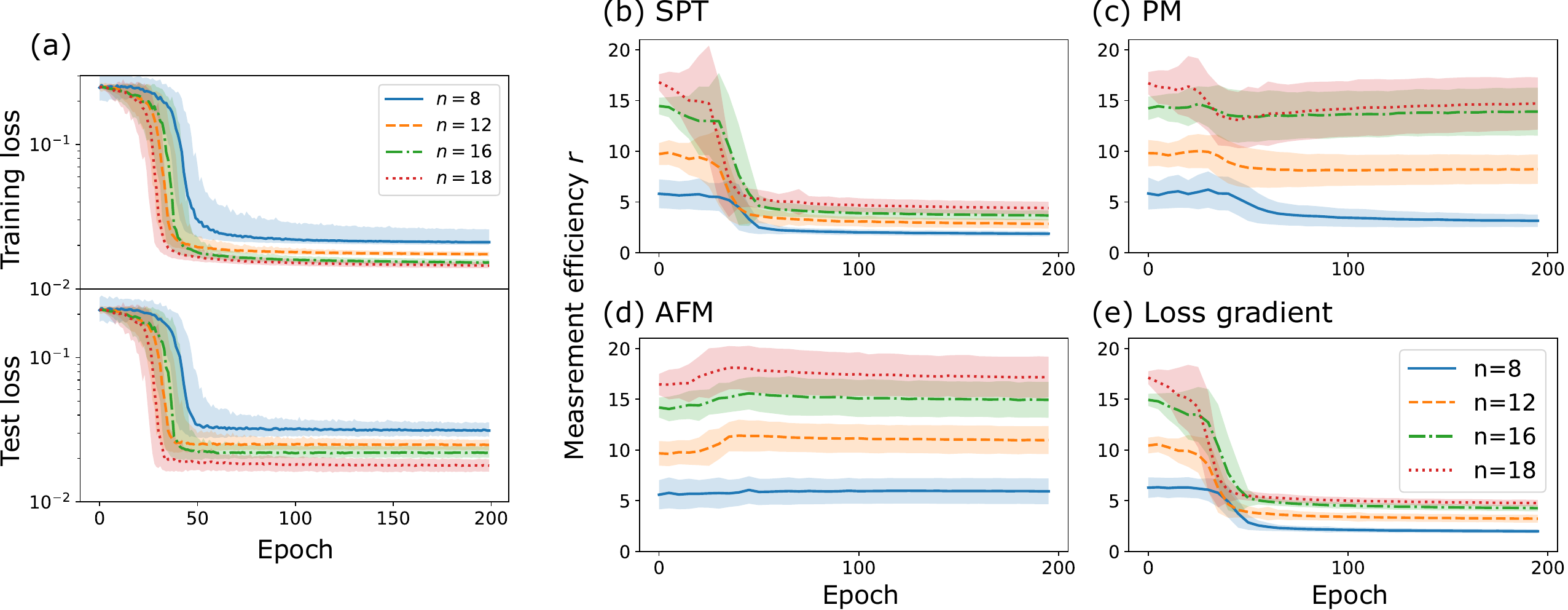}
		\caption{
			(a) Changes in training and test loss functions for several numbers of qubits $n$.
			The lines and shaded areas depict the median and the 10th–90th percentiles for 200 sets of random initial parameters, respectively.}
		(b)--(e) Changes in relative measurement efficiency during training.
		(b)--(d) show the efficiency $r$ for different inputs, SPT, PM, and AFM states,
		whereas (e) shows the efficiency for measuring the loss gradient by the first parameter.
		At each epoch, we simulate experiments with 1000 shots 10000 times, estimate $\sigma_0$ and $\sigma_\text{sp}$, and calculate the efficiency $r=(\sigma_0/\sigma_\text{sp})^2$.
		The solid lines and shaded areas depict the mean values and standard deviations, respectively, for 200 sets of initial parameters. 
		Except for evaluating the efficiency, we optimize the circuit using the exact expectation value (i.e., without statistical errors).
		\label{fig: 4eff}
	\end{figure*}
	%%%%%%%%%%%%%%%%%%%%%%%%%%%%%%%%%%%%%%%%%%%%%%%%%%%%%%%%%%%%%%%%%%%%%%%%

	%%%%%%%%%%%%%%%%%%%%%%%%%%%%%%%%%%%%%%%%%%%%%%%%%%%%%%%%%%%%%%%%%%%%%%%%
	\begin{figure*}[t]
		\centering
		\includegraphics[width=\linewidth]{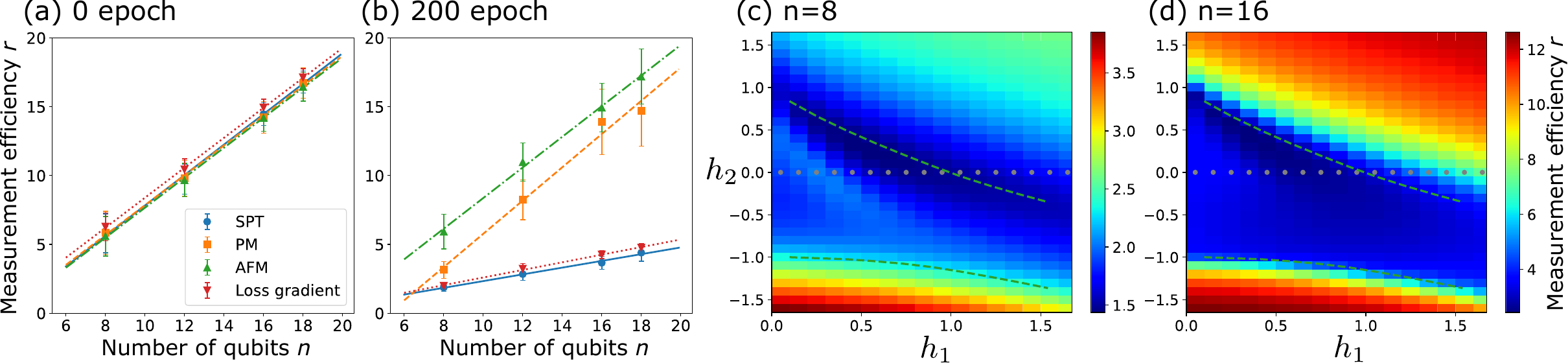}
		\caption{
			(a), (b) Relative measurement efficiency $r$ with varying the number of qubits $n$ for SPT, PM, AFM, and loss gradient at (a) 0 and (b) 200 epoch in Fig.~\ref{fig: 4eff}.
			The four straight lines fit the corresponding types of data points. 
			The error bars denote the standard deviations for 200 sets of initial parameters. 
			(c), (d) Relative measurement efficiency $r$ on $h_1$-$h_2$ plane after 200 epochs for (c) $n=8$ and (d) 16 qubits.
			The color denotes the magnitude of the relative measurement efficiency.
			The gray dots and dashed green lines denote our training data and phase boundaries computed by DMRG, respectively.
		}
		\label{fig: 4eff_2}
	\end{figure*}
	%%%%%%%%%%%%%%%%%%%%%%%%%%%%%%%%%%%%%%%%%%%%%%%%%%%%%%%%%%%%%%%%%%%%%%%%

	%%%%%%%%%%%%%%%%%%%%%%%%%%%%%%%%%%%%%%%%%%%%%%%%%%%%%%%%%%%%%%%%%%%%%%%%
	\subsection{Quantification of measurement efficiency} \label{sec: eff}
	
	The previous subsection reveals that the sp-QCNN suppresses the statistical errors in estimating the gradient and accelerates the learning process due to its high measurement efficiency.
	Here, we numerically quantify the measurement efficiency of sp-QCNN, showing that it improves the measurement efficiency by a factor of $\mO(n)$ in the quantum phase recognition task.
	The measurement efficiency is quantified by the ratio of the variances in the nonsplitting and sp-QCNNs, as shown in Fig.~\ref{fig: variance} and Eq.~\eqref{eq: r} (we assume that the nonsplitting QCNN consists of the same unitary $V_i$ as the sp-QCNN).
	We calculate the measurement efficiency for $n=8,12,16$ and $18$ to identify the scaling with respect to the number of qubits.
	In this verification, the changes in the measurement efficiency are computed during the training process for three typical input states: SPT, PM, and AFM states, which are the eigenstates of $H$ for $(h_1,h_2)=(0,0), (+\infty,0),$ and $(0,-\infty)$.
	We also explore the efficiency of measuring the loss gradient for the first parameter.
	%To clarify the scaling of the measurement efficiency, we compute the efficiency for $n=8,12,16$ and $18$.
	The unitary circuit consists of the translationally symmetric ansatz~\eqref{eq: ansatz} with $d=10$ and is split such that the number of qubits in a branch varies as $8\to4\to2\to1$, $12\to6\to3\to1$, $16\to8\to4\to2\to1$, and $18\to9\to3\to1$ for $n=8,12,16$, and $18$, respectively.
	Also, to distinguish between the effects of statistical errors on the circuit optimization and the observable measurement, we train the variational circuit with exact gradient (i.e., without statistical errors in estimating gradient) and estimate the measurement efficiency with a finite shot simulation at each epoch. 
	Figure~\ref{fig: 4eff}(a) shows the training and test losses in this setup, which indicates good trainability and generalization up to $n=18$.

	Figures~\ref{fig: 4eff}(b)--(e) show the changes in the relative measurement efficiency $r$ during training for the three inputs and loss gradient.
	For the PM and AFM states [(c) and (d)], the efficiency $r$ is high at the beginning and does not significantly decrease during training.
	For the SPT state and loss gradient [(b) and (e)], $r$ is initially high but decreases as training, finally converging to a small value ($r=2$--$5$).
	These results imply that the improvement rate of measurement efficiency strongly depends on the input data, what we measure, and the stage of learning.
	Even for the SPT state and loss gradient, the final efficiency is higher than one, indicating that the measurement in the sp-QCNN is more efficient than that in the nonsplitting QCNN.

	Figures~\ref{fig: 4eff_2}(a) and (b) show the relative measurement efficiency with varying the number of qubits $n$ at 0 and 200 epochs for the four cases (cf. Fig.~\ref{fig: 4eff}).
	At 0 epoch (a), all the data points are nearly aligned on a straight line.
	This result supports that measurement efficiency is improved by a factor of $\mO(n)$ in the early stage of learning and is consistent with the previous argument based on randomness.
	Even at 200 epoch (b), we can fit the data points in straight lines within their error bars, suggesting that the efficiency is also improved by a factor of $\mO(n)$ in the final stage of learning.
	In other words, compared with the nonsplitting QCNN, the sp-QCNN can reduce the number of shots required to achieve a certain estimation accuracy of expectation values by a factor of $\mO(1/n)$ throughout the learning process.

	We also investigate the measurement efficiency for predicting the phase diagram by the trained sp-QCNN with $n=8$ and $16$ qubits [Figs.~\ref{fig: 4eff_2}(c) and (d)].
	By comparing these figures, we notice that the efficiency $r$ for $n=16$ is more than twice that for $n=8$ in most areas.
	This result implies the $\mO(n)$ times improvement for prediction.
	We also observe that the efficiency is low in the SPT phase but relatively high in the PM and AFM phases, a trend evident in Fig.~\ref{fig: 4eff} as well.
	We infer that this phenomenon is due to the following reason.
	For the SPT state, the expectation value of $\sum_j Z_j/n$ after training is almost one because we have assigned the label as $y_i=1$ for the SPT phase in the loss function, which means that $U\ket{\phi_\text{SPT}} \sim \ket{0\cdots0}$.
	Given that the sp-QCNN has no advantages for measuring $\ket{0\cdots0}$, the measurement efficiency is not significantly improved.
	For complete understanding, additional analyses must be conducted in future works.

	%%%%%%%%%%%%%%%%%%%%%%%%%%%%%%%%%%%%%%%%%%%%%%%%%%%%%%%%%%%%%%%%%%%%%%%%
	%%%%%%%%%%%%%%%%%%%%%%%%%%%%%%%%%%%%%%%%%%%%%%%%%%%%%%%%%%%%%%%%%%%%%%%%
	%%%%%%%%%%%%%%%%%%%%%%%%%%%%%%%%%%%%%%%%%%%%%%%%%%%%%%%%%%%%%%%%%%%%%%%%
	\section{Conclusions} \label{sec: conclusion}
	
	In this study, we have proposed a new QNN architecture, sp-QCNN, which reduces measurement costs by exploiting the translational symmetry of data as prior knowledge.
	In the sp-QCNN, we symmetrize and split the QCNN circuit to parallelize the computation, thus improving the measurement efficiency.
	We have demonstrated the advantage of the sp-QCNN for the quantum phase recognition task: it has high classification performance for this task and can improve the measurement efficiency by a factor of $\mO(n)$.
	In a realistic setting where measurement resources are limited, the sp-QCNN can enhance the speed and stability of the learning process.
	These results present a new possibility for the symmetry-based architecture design of QNN and bring us one step closer to achieving the quantum advantages of the QCNN in near-term quantum devices.

	This work offers some research directions for the future. 
	First, finding practical applications of the sp-QCNN is crucial for quantum advantages.
	A promising candidate is the research of solids, where the sp-QCNN could offer some hints on unsolved problems in condensed matter physics, such as the phase diagrams of the Hubbard model~\cite{Arovas2022-ed} and the kagome antiferromagnetic Heisenberg model~\cite{Yan2011-df}.
	The second direction is further studies of symmetry-based architecture design to reduce measurement costs.
	Although this work has provided a new approach for QML, its coverage is limited to data with translational symmetry.
	Hence, generalization to other symmetries, such as the space group, is intriguing and fruitful and may be applied to chemical molecules as well as solid-state materials.
	The third direction is to find a better ansatz.
	Although this work establishes the basis of the sp-QCNN, the best $V_i$ for a given problem remains unclear.
	In general, low expressivity tends to result in poor QML accuracy, while excessively high expressivity can lead to barren plateaus~\cite{Anschuetz2022-gb, Holmes2022-uk}.
	Therefore, finding an ansatz with appropriate expressivity depending on a problem is helpful to realize sp-QCNN experimentally.

	Finally, we provide several open issues on the sp-QCNN.
	This work has shown that the sp-QCNN has sufficient expressivity, trainability, generalization, and measurement efficiency to solve the phase recognition task.
	However, whether it can solve other complicated tasks remains unclear.
	In particular, the translational symmetry of $V_i$ could suppress expressivity and limit solvable tasks in general.
	Uncovering the possibilities and limitations of the sp-QCNN is an important open issue. 
	For trainability, elucidating whether barren plateaus exist in the sp-QCNN is crucial.
	In the conventional QCNN, barren plateaus do not appear due to its unique architecture: the logarithmic circuit depth and the locality of unitary operations and observables~\cite{Pesah2021-py, McClean2018-qf, Cerezo2021-tq, Ortiz_Marrero2021-ni}.
	Considering that the sp-QCNN shares these properties with the conventional QCNN, we suggest that no barren plateaus will appear even in the sp-QCNN~\cite{Tuysuz2022-zi}.
	The results in this paper show that the training of the sp-QCNN works well up to $n=18$ qubits, supporting our conjecture.
	The analysis of measurement efficiency is also an important research issue.
	Besides, while we observed the $\mO(n)$ times improvement of measurement efficiency in the phase recognition task of Hamiltonian~\eqref{eq: Ham}, it remains unclear whether the measurement efficiency is improved by a factor of $\mO(n)$ in other problems as well.
	More thorough analyses are necessary for complete verification.

	\section*{Acknowlegments}
	Fruitful discussions with Masatoshi Ishii, Tomochika Kurita, Yuichi Kamata, and Yasuhiro Endo are gratefully acknowledged.

	%%%%%%%%%%%%%%%%%%%%%%%%%%%%%%%%%%%%%%%%%%%%%%%%%%%%%%%%%%%%%%%%%%%%%%%%%%%%%%%%%%%%%%%%%%%%%%%%%%%%%%%
	%%%%%%%%%%%%%%%%%%%%%%%%%%%%%%%%%%%%%%%%%%%%%%%%%%%%%%%%%%%%%%%%%%%%%%%%%%%%%%%%%%%%%%%%%%%%%%%%%%%%%%%
	%%%%%%%%%%%%%%%%%%%%%%%%%%%%%%%%%%%%%%%%%%%%%%%%%%%%%%%%%%%%%%%%%%%%%%%%%%%%%%%%%%%%%%%%%%%%%%%%%%%%%%%
	\appendix

	%%%%%%%%%%%%%%%%%%%%%%%%%%%%%%%%%%%%%%%%%%%%%%%%%%%%%%%%%%%%%%%%%%%%%%%%%%%%%%%%%%%%%%%%%%%%%%%%%%%%%%%
	\section{Implementation of circuit splitting} \label{sec: swap}
	
	%%%%%%%%%%%%%%%%%%%%%%%%%%%%%%%%%%%%%%%%%%%%%%%%%%%%%%%%%%%%%%%%%%%%%%%%
	\begin{figure}[t]
		\centering
		\includegraphics[width=\linewidth]{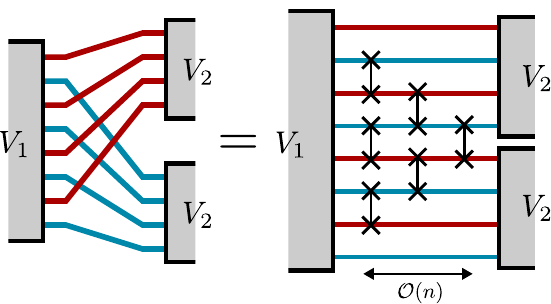}
		\caption{
			An implementation of circuit splitting using local SWAP gates.
			This figure shows the circuit splitting of $n$ qubits to two branches ($n$ even).
			The SWAP procedure consists of $(n-2)/2$ steps.
			For step $d$, we swap $j$th and ($j+1$)th qubits for $j=d+1,d+3,\cdots,n-d-1$.
			This circuit uses $(n^2-2n)/8$ SWAP gates with a depth of $(n-2)/2$.
		}
		\label{fig: swap}
	\end{figure}
	%%%%%%%%%%%%%%%%%%%%%%%%%%%%%%%%%%%%%%%%%%%%%%%%%%%%%%%%%%%%%%%%%%%%%%%%

	In real hardware, the implementation of circuit splitting depends on the hardware connectivity.
	For example, in quantum computers with all-to-all connectivity such as ion trap devices, special processes for circuit splitting are not necessary because the unitary operations on each branch after splitting can be implemented with the qubits being separated.
	Conversely, in quantum computers where only local entangling operations on neighboring qubits are available, we have to swap the qubits in order to implement the unitary operations after splitting.
	For concreteness, let us consider a case where $n$ qubits are split into two branches (we assume that $n$ is even).
	Figure~\ref{fig: swap} illustrates the SWAP circuit for splitting, which uses $\mO(n^2)$ SWAP gates with a depth of $\mO(n)$.
	Although conventional QCNNs also require SWAP gates to rearrange the remaining qubits in the pooling layers, our sp-QCNN needs more SWAP gates in general.

	%%%%%%%%%%%%%%%%%%%%%%%%%%%%%%%%%%%%%%%%%%%%%%%%%%%%%%%%%%%%%%%%%%%%%%%%%%%%%%%%%%%%%%%%%%%%%%%%%%%%%%%
	\section{The number of eigenstates of $Z_\text{avg}$} \label{sec: DOS}

	Here, we derive Eq.~\eqref{eq: Dn}, where the number of eigenstates of $Z_\text{avg}$ for translationally symmetric states is
	\begin{align}
		D_n(s) \sim \frac{1}{(1+s^2)^{n/2}}
	\end{align}
	with an eigenvalue $s$.
	For convenience, we consider $Z_\text{tot}=\sum_j Z_j$ rather than $Z_\text{avg}=Z_\text{tot}/n$.
	In the sp-QCNN, we measure $Z_\text{tot}$ whose eigenvalues are $\pm n, \pm (n-2),\cdots$ and obtain one of the eigenvalues every shot.
	In addition, the output state is translationally symmetric in the sp-QCNN.
	Hence, for simplicity, we now focus on the $T$-invariant eigenspace of $Z_\text{tot}$ with an eigenvalue $z$, $V_z$ (i.e., $T\ket{\phi}=\ket{\phi}$ and $Z_\text{tot}\ket{\phi}=z\ket{\phi}$ for any $\ket{\phi}\in V_z$).
	Below, we investigate the dimension of $V_z$.

	To this end, we introduce a cyclic group generated by $T$,
	\begin{align}
		G_n=\{I, T, T^2,\cdots,T^{n-1}\}.
	\end{align}
	Let $M_z$ be the set of the eigenstates of $Z_\text{tot}$ with an eigenvalue $z$ in the computational basis (e.g., $M_{n-2}=\{\ket{10\cdots00}, \cdots, \ket{00\cdots01}\}$).
	Then, we define an equivalence relation ${\sim}$ by $G_n$ in $M_z$:
	for $\ket{a}, \ket{b} \in M_z$, $\ket{a} \sim \ket{b}$ holds if and only if $g\ket{a}=\ket{b}$ with $^\exists g\in G_n$.
	We also define the equivalence class of $\ket{a}\in M_z$ as $[a]=\{\ket{x}\in M_z \,|\, \ket{x} \sim \ket{a}\}$ and the quotient set as $M_z/G_n=\{[a] \,|\, \ket{a}\in M_z\}$.
	The elements of $M_z/G_n$ correspond one-to-one to the bases of $V_z$, such that $\ket{\Psi_i} = \sum_{\ket{\phi} \in [\Phi_i]} \ket{\phi} / \mathcal{N}$, where $\ket{\Psi_i}$ is the base of $V_z$, $ [\Phi_i]$ is the element of $M_z/G_n$, and $\mathcal{N}$ is the normalization factor ($T\ket{\Psi_i}=\ket{\Psi_i}$ can be easily checked).
	Therefore, $\text{dim}V_z=|M_z/G_n|$ holds, where $|A|$ is the number of elements in $A$. 
	Using Burnside's lemma~\cite{Rotman1994-ez}, we have $\text{dim}V_z$ as follows:
	\begin{align}
		\text{dim}V_z = |M_z/G_n| = \frac{1}{|G_n|} \sum_{g\in G_n} |M_z^g|, \label{eq: Burnside}
	\end{align}
	where $M_z^g = \{\ket{\phi}\in M_z \,\,|\,\, g \ket{\phi}=\ket{\phi}\}$.
	Then, the following theorem holds.
	\begin{thm}
		For $z\neq \pm n$, the following relation holds in the limit of $n\to\infty$:
		\begin{align}
			F_z \equiv \left. \text{dim}V_z \middle/ \frac{1}{n}\binom{n}{\ell_z} \right. \xrightarrow{n\to\infty} 1
		\end{align}
		with $\ell_z=(n+z)/2$. 
		Here $\binom{\cdot}{\cdot}$ denotes the binomial coefficient.
		This theorem states that the asymptotic form of $\text{dim}V_z$ is $\binom{n}{\ell_z}/n$.
	\end{thm}

	%%%%%%%%%%%%%%%%%%%%%%%%%%%%%%%%%%%%%%%%%%%%%%%%%%%%%%%%%%%%%%%%%%%%%%%%
	\begin{figure}[t]
		\centering
		\includegraphics[width=\linewidth]{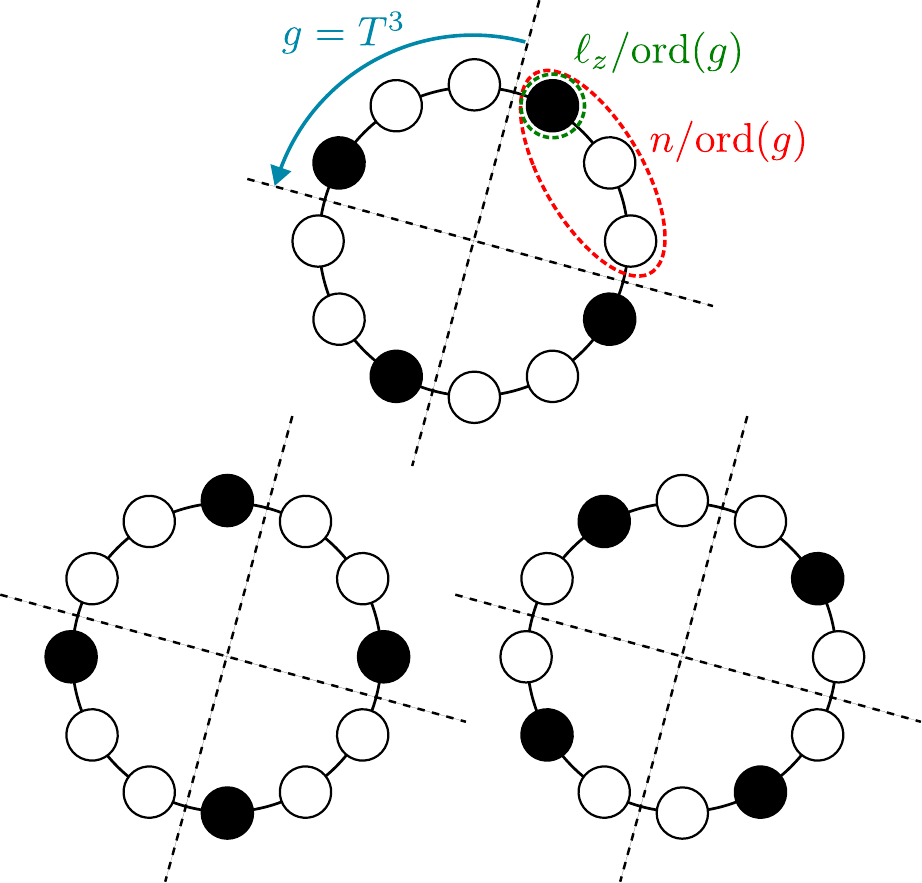}
		\caption{
			Illustration for calculating $|M_z^g|$ with $n=12, \ell_z=4$, and $g=T^3$.
			Each white (black) circle indicates a single qubit state of $\ket{0}$ ($\ket{1}$), and $n=12$ and $\ell_z=4$ mean that there are eight ($=n-\ell_z$) white and four ($=\ell_z$) black circles in total.
			Given that the order of $g$ is four ($\because g^4=I$), we first divide the qubits into four sets, each consisting of three ($=n/\text{ord}(g)$) qubits.
			In all the sets, the configuration of white and black circles must be the same as each other because of the condition that $g$ does not change the state. 
			Therefore, each set has two white and one ($=\ell_z/\text{ord}(g)$) black circles, and there are three ($=\binom{n/\text{ord}(g)}{\ell_z/\text{ord}(g)}$) possible configurations shown in the figure.
		}
		\label{fig: cyclic}
	\end{figure}
	%%%%%%%%%%%%%%%%%%%%%%%%%%%%%%%%%%%%%%%%%%%%%%%%%%%%%%%%%%%%%%%%%%%%%%%%

	\begin{proof}
		
		Since $F_z = F_{-z}$ trivially holds, we focus on $-n+2 \leq z \leq0$, or, $1\leq \ell_z \leq \lfloor n/2 \rfloor$ ($\lfloor \cdot\rfloor$ is the floor function). 
		We first rewrite Eq.~\eqref{eq: Burnside} as
		\begin{align}
			\text{dim}V_z
			&= \frac{M_z^I}{|G_n|} + \frac{1}{|G_n|} \sum_{g\in G_n \setminus \{I\}} |M_z^g| \notag \\
			&= \frac{1}{n}\binom{n}{\ell_z} + \frac{1}{n} \sum_{g\in G_n \setminus \{I\}} |M_z^g|.
		\end{align}
		Therefore, $F_z$ is reduced to 
		\begin{align}
			F_z = 1 + \left. \sum_{g\in G_n \setminus \{I\}} |M_z^g| \middle/ \binom{n}{\ell_z} \right.. \label{eq: Fz}
		\end{align}
		We will evaluate the second term in this equation.

		To calculate $|M_z^g|$, we define the order of $g\in G_n$, $\text{ord}(g)$, as the number of elements in the subgroup generated by $g$ (i.e., $\{g^0,g^1,g^2,\cdots,g^{k-1}\}$ with $g^k=I$).
		Note that $\text{ord}(g)$ is a divisor of $n$.
		Thereby, $|M_z^g|$ is written as follows:
		\begin{align}
			|M_z^g| =
			\begin{dcases}
				0 & \ell_z/\text{ord}(g) \notin \mathbb{Z} \\
				\binom{n/\text{ord}(g)}{\ell_z/\text{ord}(g)} & \ell_z/\text{ord}(g) \in \mathbb{Z}. \label{eq: Mzg}
			\end{dcases}
		\end{align}
		Figure~\ref{fig: cyclic} shows a graphical description of Eq.~\eqref{eq: Mzg} as an example for $n=12, \ell_z=4$, and $g=T^3$.

		Based on Eq.~\eqref{eq: Mzg}, one can straightforwardly show that the second term in Eq.~\eqref{eq: Fz} vanishes in the limit of $n\to\infty$ for $\ell_z = 1,2$ by noticing that $\text{ord}(g) = 1$ only for $g=I$ and $\text{ord}(g) = 2$ only for $g=T^{n/2}$.
		Thus, we focus on $3 \leq \ell_z\leq \lfloor n/2 \rfloor$.
		Because of $\text{ord}(g) \geq 2$ for $g \neq I$, we have
		\begin{align}
			|M_z^g| \leq \binom{\lfloor n/2 \rfloor}{\lfloor \ell_z/2 \rfloor}. \label{eq: ineqM}
		\end{align}
		This inequality can be shown by considering the properties of binomial coefficients: $\binom{a}{b} < \binom{a'}{b}$ ($a<a'$) and $\binom{a}{0} < \binom{a}{1}<\cdots<\binom{a}{\lfloor a/2 \rfloor}$ (note that $3\leq \ell_z \leq \lfloor n/2 \rfloor$).
		Using Eq.~\eqref{eq: ineqM}, the second term in Eq.~\eqref{eq: Fz} is bounded as follows:
		\begin{align}
			0\leq \left. \sum_{g\in G_n \setminus \{I\}} |M_z^g| \middle/ \binom{n}{\ell_z} \right. \leq \left. n\binom{\lfloor n/2 \rfloor}{\lfloor \ell_z/2 \rfloor} \middle/ \binom{n}{\ell_z} \right. \equiv A_{\ell_z}.
		\end{align}
		The right-hand side of this inequality, $A_{\ell_z}$, approaches zero in $n\to\infty$ for $3 \leq \ell_z \leq \lfloor n/2 \rfloor$, which can be proven by showing $A_3 \xrightarrow{n\to\infty} 0$ and $0 < A_{\lfloor n/2 \rfloor} < \cdots <A_4<A_3$ by definition of $A_{\ell_z}$.
		Therefore, we have 
		\begin{align}
			\left. \sum_{g\in G_n \setminus \{I\}} |M_z^g| \middle/ \binom{n}{\ell_z} \right. \xrightarrow{n\to\infty} 0. \label{eq: Mzlimit}
		\end{align}
		As mentioned above, this limit holds true even for $\ell_z=1,2$.
		Substituting Eq.~\eqref{eq: Mzlimit} to Eq.~\eqref{eq: Fz}, we obtain
		\begin{align}
			F_z \xrightarrow{n\to\infty} 1.
		\end{align}
		with $\ell_z \neq 0, n$.

	\end{proof}

	This theorem states that $\text{dim}V_z$ asymptotically approaches $\binom{n}{\ell_z}/n$ (except for $\text{dim}V_n=\text{dim}V_{-n}=1$).
	Using Stirling's formula ($n! \sim \sqrt{2\pi n}(n/e)^n$), we have
	\begin{align}
		\text{dim}V_z &\sim 2^n\sqrt{\frac{2}{\pi n^3}} D_n(s),
	\end{align}
	where we have defined 
	\begin{align}
		D_n(s) \equiv \left[ (1+s)^{1+s+\frac{1}{n}} (1-s)^{1-s+\frac{1}{n}}\right]^{-n/2} 
	\end{align}
	with $s=z/n$.
	For large $n$, $D_n(s)$ rapidly decreases to vanish away from the origin.
	Therefore, we expand the denominator of $D_n(s)$ in $s$, obtaining
	\begin{align}
		D_n(s) 
		&= \Big( 1+(1+\mO(1/n))s^2 + \mO(s^4) \Big)^{-n/2} \notag \\
		&\sim \frac{1}{(1+s^2)^{n/2}},
	\end{align}
	for sufficiently small $s$.
	The width of $D_n(s)$ is $\mO(1/\sqrt{n})$ in $n\to\infty$, leading to the $\mO(n)$ times improvement of measurement efficiency in the sp-QCNN (see Sec.~\ref{sec: random}).
	Finally, we remark that this discussion is approximately valid for large but finite $n$ while this appendix considers the limit of $n\to\infty$.
	In fact, in Sec.~\ref{sec: eff}, we have observed the clear $\mO(n)$ scaling for $n=18$ at the beginning of training where the output state is almost random.

	%\bibliography{refs_v3}
	%bibファイルのlanguageを消さないとダメ！
	

\end{document}